\documentclass[longauth]{aa}

\usepackage{graphicx}
\usepackage{txfonts}
\usepackage{natbib}
\usepackage{amsmath}
\usepackage{booktabs}
\usepackage{array}
\usepackage{siunitx}
\usepackage[colorlinks=true,allcolors=blue]{hyperref}

\newcommand{\ffarcs}{\mbox{\ensuremath{.\!\!^{\prime\prime}}}}

\newcommand{\ffsecond}{\mbox{\ensuremath{.\!\!^{\rm s}}}}

\newcolumntype{L}[1]{>{\raggedright\let\newline\\\arraybackslash\hspace{0pt}}m{#1}}
\newcolumntype{C}[1]{>{\centering\let\newline\\\arraybackslash\hspace{0pt}}m{#1}}

\bibpunct{(}{)}{;}{a}{}{,}

\begin{document}

\title{Shadows cast on the transition disk of HD~135344B
\thanks{Based on observations collected at the European Southern Observatory, Chile, ESO No. 095.C-0273(A) and 095.C-0273(D).}}
\subtitle{Multiwavelength VLT/SPHERE polarimetric differential imaging}

\author{
T.~Stolker\inst{1}
\and C.~Dominik\inst{1}
\and H.~Avenhaus\inst{2,21}
\and M.~Min\inst{3,1}
\and J.~de Boer\inst{4,5}
\and C.~Ginski\inst{4}
\and H.~M.~Schmid\inst{6}
\and A.~Juhasz\inst{7}
\and A.~Bazzon\inst{6}
\and L.~B.~F.~M.~Waters\inst{3,1}
\and A.~Garufi\inst{6}
\and J.-C.~Augereau\inst{8,9}
\and M.~Benisty\inst{8,9}
\and A.~Boccaletti\inst{10}
\and Th.~Henning\inst{11}
\and M.~Langlois\inst{13,14}
\and A.-L.~Maire\inst{11}
\and F.~M\'{e}nard\inst{12,2}
\and M.~R.~Meyer\inst{6}
\and C.~Pinte\inst{12,2}
\and S.~P.~Quanz\inst{6}
\and C.~Thalmann\inst{6}
\and J.-L.~Beuzit\inst{8,9}
\and M.~Carbillet\inst{15}
\and A.~Costille\inst{14}
\and K.~Dohlen\inst{14}
\and M.~Feldt\inst{11}
\and D.~Gisler\inst{6}
\and D.~Mouillet\inst{8,9}
\and A.~Pavlov\inst{11}
\and D.~Perret\inst{10}
\and C.~Petit\inst{16}
\and J.~Pragt\inst{17}
\and S.~Rochat\inst{8,9}
\and R.~Roelfsema\inst{17}
\and B.~Salasnich\inst{18}
\and C.~Soenke\inst{19}
\and F.~Wildi\inst{20}
}

\institute{
Anton Pannekoek Institute for Astronomy, University of Amsterdam, Science Park 904, 1098 XH Amsterdam, The Netherlands\\
\email{T.Stolker@uva.nl}
\and Departamento de Astronom\'{i}a, Universidad de Chile, Casilla 36-D, Santiago, Chile
\and SRON Netherlands Institute for Space Research, Sorbonnelaan 2, 3584 CA Utrecht, The Netherlands
\and Leiden Observatory, Leiden University, PO Box 9513, 2300 RA Leiden, The Netherlands
\and ESO, Alonso de C\'{o}rdova 3107, Vitacura, Casilla 19001, Santiago de Chile, Chile
\and Institute for Astronomy, ETH Zurich, Wolfgang-Pauli-Strasse 27, 8093 Zurich, Switzerland
\and Institute of Astronomy, Madingley Road, Cambridge CB3 OHA, UK
\and Universit\'{e} Grenoble Alpes, IPAG, F-38000 Grenoble, France
\and CNRS, IPAG, F-38000 Grenoble, France
\and LESIA, Observatoire de Paris, CNRS, Universit\'{e} Paris Diderot, Universit\'{e} Pierre et Marie Curie, 5 place Jules Janssen, 92190 Meudon, France
\and Max Planck Institute for Astronomy, K\"{o}nigstuhl 17, D-69117 Heidelberg, Germany
\and UMI-FCA, CNRS/INSU, France (UMI 3386)
\and CRAL, UMR 5574, CNRS, Universit\'{e} Lyon 1, 9 avenue Charles Andr\'{e}, 69561 Saint Genis Laval Cedex, France
\and Aix Marseille Universit\'{e}, CNRS, LAM (Laboratoire d'Astrophysique de Marseille) UMR 7326, 13388, Marseille, France
\and Universit\'{e} Nice-Sophia Antipolis, CNRS, Observatoire de la Côte d'Azur, Laboratoire J.-L. Lagrange, CS 34229, 06304 Nice cedex 4, France
\and ONERA - Optics Department, 29 avenue de la Division Leclerc, F-92322 Chatillon Cedex, France
\and NOVA Optical-Infrared Instrumentation Group at ASTRON, Oude Hoogeveensedijk 4, 7991 PD Dwingeloo, The Netherlands
\and INAF-Osservatorio Astronomico di Padova, Vicolo dell'Osservatorio 5, 35122, Padova, Italy
\and European Southern Observatory, Karl-Schwarzschild-Strasse 2, D-85748 Garching, Germany
\and Geneva Observatory, University of Geneva, Chemin des Mailettes 51, 1290 Versoix, Switzerland
\and Millennium Nucleus "Protoplanetary Disks", Chile
}

\date{Received ?; accepted ?}

\abstract {The protoplanetary disk around the F-type star HD~135344B (SAO~206462) is in a transition stage and shows many intriguing structures both in scattered light and thermal (sub-)millimeter emission which are possibly related to planet formation processes.}
{We study the morphology and surface brightness of the disk in scattered light to gain insight into the innermost disk regions, the formation of protoplanets, planet-disk interactions traced in the surface and midplane layers, and the dust grain properties of the disk surface.}
{We have carried out high-contrast polarimetric differential imaging (PDI) observations with \mbox{VLT/SPHERE} and obtained polarized scattered light images with \mbox{ZIMPOL} in $R$- and $I$-band and with \mbox{IRDIS} in $Y$- and $J$-band. The scattered light images and surface brightness profiles are used to study in detail structures in the disk surface and brightness variations. We have constructed a 3D radiative transfer model to support the interpretation of several detected shadow features.}
{The scattered light images reveal with unprecedented angular resolution and sensitivity the spiral arms as well as the 25~au cavity of the disk. Multiple shadow features are discovered on the outer disk with one shadow only being present during the second observation epoch. A positive surface brightness gradient is observed in the stellar irradiation corrected ($r^2$-scaled) images in southwest direction possibly due to an azimuthally asymmetric perturbation of the temperature and/or surface density by the passing spiral arms. The disk integrated polarized flux, normalized to the stellar flux, shows a positive trend towards longer wavelengths which we attribute to large ($2\pi a\gtrsim\lambda$) aggregate dust grains in the disk surface. Part of the the non-azimuthal polarization signal in the $U_\phi$ image of the $J$-band observation could be the result of multiple scattering in the disk.}
{The detected shadow features and their possible variability have the potential to provide insight into the structure of and processes occurring in the innermost disk regions. Possible explanations for the presence of the shadows include a $22\degr$ misaligned inner disk, a warped disk region that connects the inner disk with the outer disk, and variable or transient phenomena such as a perturbation of the inner disk or asymmetric accretion flows. The spiral arms are best explained by one or multiple protoplanets in the exterior of the disk although no gap is detected beyond the spiral arms up to 1\ffarcs0.}

\keywords{Stars: individual: HD~135344B (SAO~206462) -- Protoplanetary disks -- Planet-disk interactions -- Methods: observational -- Instrumentation: high angular resolution -- Techniques: polarimetric}

\maketitle

\section{Introduction}\label{sec:introduction}

Protoplanetary disks around young stars are the birth environments of planets and studying them is essential for understanding the planet formation process. High-resolution imaging of protoplanetary disks in scattered light and thermal (sub-)millimeter emission reveals morphological structures related to disk evolution and planet-disk interactions among others. Transition disks with large dust cavities are of particular interest because they comprise a small sample of protoplanetary disks that are thought to be in the last stage of disk evolution and possibly planet formation during which the disk evolves from gas-rich protoplanetary disk to gas-poor debris disk. The spectral energy distributions (SED) of transition disks are characterized by a deficit in the infrared excess, relative to typical primordial disk targets, which is usually interpreted as a consequence of dust depletion in the inner regions of the disk \citep[e.g.,][]{strom1989,skrutskie1990,espaillat2014}. Imaging of transition disks can unambiguously establish whether a dust cavity is indeed present as well as other structures in the disk. Scattered light imaging at optical and near-infrared wavelengths traces small (micron-sized) dust grains in the disk surface whereas resolved (sub-)millimeter continuum observations trace the thermal radiation of larger (mm-sized) dust grains in the disk midplane.
 
HD~135344B (SAO~206462) is an \mbox{F4Ve star} \citep{dunkin1997} with a transition disk which is located in the \mbox{Scorpius OB2-3} (Upper Centaurus Lupus) star forming region at a distance of $140\pm42$~pc with an estimated age of $8\substack{+8\\-4}$~Myr \citep{vanboekel2005}. The star is part of a visual binary system and has a separation of $21\arcsec$ from the primary star, HD~135344 (SAO~206463) \citep{mason2001}, which is an \mbox{A0V star} \citep{houk1982}. HD~135344B has been classified as a group~Ib source by \citet{meeus2001} because the SED rises in the mid-infrared yet lacks a \SI{10}{\micro\meter} silicate emission feature. Group~I sources are expected to feature a large gap and exhibit mid-infrared excess dominated by reprocessing of stellar flux by the rim and the flared surface of the outer disk \citep{maaskant2013}. The infrared excess of the HD~135344B disk is ${\rm L}_{\rm IR}/{\rm L}_* = 0.53$ which is a typical value for group~I protoplanetary disks \citep{dominik2003}. The main parameters of the system are summarized in Table~\ref{table:main_parameters}.

Analysis of the SED by \citet{brown2007} suggested that a large dust cavity with a radius of 45~au is present in the disk. Dust continuum (sub-)millimeter observations have confirmed the presence of the dust cavity and show a crescent-shaped asymmetry with a factor of two emission contrast due to an azimuthal asymmetry in the thermal disk structure and/or the distribution of the large dust grains \citep{brown2009,lyo2011,andrews2011,perez2014,pinilla2015}. \mbox{ALMA} observations of $^{13}$CO and C$^{18}$O revealed a 25~au gas cavity in which the CO surface density is reduced by a factor of 10000 \citep{vandermarel2016}.

The HD~135344B disk was first detected in scattered light with the \emph{Hubble Space Telescope} (HST) which showed part of the outer disk but there was no unambiguous detection of any substructures \citep{grady2009}. Subaru/HiCIAO $H$-band polarimetric imaging observations by \citet{muto2012} revealed two spiral arms which were discussed in the context of spiral density waves. \citet{garufi2013} carried out VLT/NACO polarimetric imaging observations in $H$- and $K_{\rm s}$-band which resolved a 28~au cavity. The outer radius of the scattered light cavity is approximately 20~au smaller than the (sub)-millimeter cavity which can be explained by spatial segregation of micron-sized and mm-sized dust grains \citep{garufi2013} which is a natural outcome of particle trapping by a pressure maximum, possibly due to planet-disk interactions \citep[e.g.,][]{rice2006,pinilla2012,zhu2012,ovelar2013}. More recently, angular differential imaging (ADI) observations with the Gemini Planet Imager (GPI) were presented by \citet{wahhaj2015} who recovered the spiral arms and new streamer-like features.

The near-infrared-excess in the SED of HD~135344B is an indicator for dust at small orbital radii \citep{brown2007,carmona2014}. A dust belt at sub-au distance from the star was confirmed by mid- and near-infrared interferometric observations \citep{carmona2014,menu2015} with possibly a misalignment of the inner and outer disk \citep{fedele2008}. Multi-epoch spectroscopic and near-infrared photometric observations show variability over the course of months possibly related to instabilities in or perturbations of the inner disk \citep{sitko2012}. HD~135344B is accreting at a rate of $10^{-8}$~M$_\odot$~yr$^{-1}$ \citep{sitko2012} and rotating near break-up velocity \citep{muller2011}.

\begin{table}
\caption{Main parameters of HD~135344B}
\label{table:main_parameters}
\centering
\bgroup
\def\arraystretch{1.1}
\begin{tabular}{l c c}
\hline\hline
Parameter & Value & Reference \\ \hline
Right ascension (J2000) & 15$^\mathrm{h}$ 15$^\mathrm{m}$ 48\ffsecond44 & (1) \\
Declination (J2000) & $-37\degr$ $09\arcmin$ 16\ffarcs059 & (1) \\
B [mag] & 9.21 $\pm$ 0.02 & (1) \\
V [mag] & 8.708 $\pm$ 0.017 & (1) \\
R [mag] & 8.302 $\pm$ 0.016 & (2) \\
I [mag] & 7.979 $\pm$ 0.017 & (2) \\
Y [mag] & 7.80 $\pm$ 0.32 & (3) \\
J [mag] & 7.279 $\pm$ 0.026 & (4) \\
H [mag] & 6.587 $\pm$ 0.031 & (4) \\
K$_{\rm s}$ [mag] & 5.843 $\pm$ 0.020 & (4) \\
A$_{\rm V}$ [mag] & 0.3 & (5) \\
Spectral type & F4Ve & (6) \\
L$_*$ [L$_\sun$]  & 7.8 & (5) \\
M$_*$ [M$_\sun$] & 1.7 & (7) \\
R$_*$ [R$_\sun$]  & 1.4 & (7) \\
T$_{\rm eff}$ [K] & 6810 & (7) \\
$\dot{\rm M}$ [M$_\sun$ yr$^{-1}$] & $(0.6-4.2) \times 10^{-8}$ & (2) \\
Distance [pc] & $140 \pm 42$ & (8) \\
Age [Myr] & $8\substack{+8 \\ -4}$ & (8) \\
Dust mass [M$_\sun$] & $1.7 \times 10^{-4}$ & (9) \\
Gas mass [M$_\sun$] & $2.4 \times 10^{-2}$ & (9) \\
Outer disk inclination [deg] & 11 & (10) \\
Outer disk position angle [deg] & 62 & (11) \\ \hline
\end{tabular}
\egroup
\tablebib{(1)~Tycho-2 Catalogue \citep{hog2000}; (2)~\citet{sitko2012}; (3)~Estimated with a weighted least squares fit of a linear function from all other magnitudes that are provided in this table; (4)~2MASS All-Sky Catalog of Point Sources \citep{skrutskie2006}; (5)~\citet{andrews2011}; (6)~\citet{dunkin1997}; (7)~\citet{muller2011}; (8)~\citet{vanboekel2005}; (9)~\citet{vandermarel2015}; (10)~\citet{lyo2011}; (11)~\citet{perez2014}.}
\end{table}

We aim to study the morphology and surface brightness of the HD~135344B disk in scattered light to gain better insight into previously detected structures. This requires high-resolution and high-contrast imaging with the best sensitivity possible. Therefore, we use polarimetric differential imaging \citep[PDI; e.g.,][]{kuhn2001,apai2004} which is a powerful technique to suppress the unpolarized speckle halo from a star and reveal the scattered light from a protoplanetary disk surface which is orders of magnitude fainter \citep[e.g.,][]{quanz2011,hashimoto2012,grady2013,avenhaus2014a,thalmann2015}.

In this paper, we present $R$-, $I$-, $Y$-, and $J$-band \mbox{VLT/SPHERE} \citep[Spectro-Polarimetric High-contrast Exoplanet REsearch;][]{beuzit2008} PDI observations of HD~135344B. The spiral arms and dust cavity are detected with unprecedented sensitivity and angular resolution and several new disk features are revealed. Total intensity images that have been obtained with \mbox{SPHERE} will be presented in a forthcoming paper (Maire et al., in prep.). In Sect.~\ref{sec:observations}, we describe the observations and data reduction procedure. In Sect.~\ref{sec:results}, we study the morphology, surface brightness and color of the scattered light. In Sect.~\ref{sec:modeling}, we use a 3D radiative transfer model to provide a possible interpretation for several detected shadow features and we fit the shape of the spiral arms in the context of planet-disk interactions. In Sect.~\ref{sec:discussion}, we discuss the results with a focus on the innermost disk regions, the formation of spiral arms, and Atacama Large Millimeter/submillimeter Array (ALMA) dust continuum observations. In Sect.~\ref{sec:conclusions}, we summarize the main conclusions.

\section{Observations and data reduction}\label{sec:observations}

The observations were carried out during the nights of 30~March~2015 ($Y$-band), 31~March~2015 ($R$- and $I$-band), and 02~May~2015 ($J$-band) with the \mbox{SPHERE} instrument which is mounted on the Nasmyth~A platform of the Unit Telescope~3 (UT3) at ESO's Very Large Telescope (VLT). \mbox{ZIMPOL} \citep[Zurich IMaging POLarimeter;][]{thalmann2008, schmid2012} and \mbox{IRDIS} \citep[Infra-Red Dual-beam Imager and Spectrograph;][]{dohlen2008,langlois2014} are the optical and near-infrared imaging component of \mbox{SPHERE}, respectively. \mbox{SPHERE} contains an extreme adaptive optics system (SAXO) to correct for atmospheric wavefront perturbations \citep{fusco2006,petit2014}.

\begin{table*}
\caption{Summary of observations}
\label{table:summary_observations}
\centering
\bgroup
\def\arraystretch{1.25}
\begin{tabular}{@{}llccccccccccc@{}}
\toprule
 & & \multicolumn{7}{c}{Integration time} & & \multicolumn{3}{c}{Observing conditions} \\
\cmidrule{3-9}\cmidrule{11-13}
Instrument \& Filter & Date & DIT\tablefootmark{a} [s] & & NDIT\tablefootmark{a} & & NINT\tablefootmark{a} & & Total\tablefootmark{a} [s] & & $\left<\mathrm{Airmass}\right>$\tablefootmark{b} & $\left<\mathrm{Seeing}\right>$\tablefootmark{c} \\ \midrule
\mbox{IRDIS} Y & 30 Mar 2015 & 0.84 & $\times$ & 30 & $\times$ & 5 & = & 126 & & 1.03 & 0\ffarcs92 (0\ffarcs07) & \\
\mbox{IRDIS} Y & 30 Mar 2015 & 0.84 & $\times$ & 15 & $\times$ & 2 & = & 25 & & 1.03 & 0\ffarcs87 (0\ffarcs07) & \\
\mbox{ZIMPOL} R & 31 Mar 2015 & 10 & $\times$ & 4 & $\times$ & 24 & = & 960 & & 1.05 & 0\ffarcs69 (0\ffarcs04) & \\
\mbox{ZIMPOL} I & 31 Mar 2015 & 10 & $\times$ & 4 & $\times$ & 24  & = & 960 & & 1.05 & 0\ffarcs69 (0\ffarcs04) & \\
\mbox{ZIMPOL} R (ND)\tablefootmark{d} & 31 Mar 2015 & 10 & $\times$ & 4 & $\times$ & 1 & = & 40 & & 1.10 & 0\ffarcs75 (0\ffarcs03) & \\
\mbox{ZIMPOL} I (ND)\tablefootmark{d} & 31 Mar 2015 & 10 & $\times$ & 4 & $\times$ & 1 & = & 40 & & 1.10 & 0\ffarcs75 (0\ffarcs03) & \\
\mbox{IRDIS} J & 02 May 2015 & 32 & $\times$ & 3 & $\times$ & 12 & = & 1152 & & 1.22 & 0\ffarcs69 (0\ffarcs07) & \\
\mbox{IRDIS} J (ND)\tablefootmark{d} & 02 May 2015 & 0.84 & $\times$ & 2 & $\times$ & 2 & = & 3& & 1.25 & 0\ffarcs69 (0\ffarcs04) & \\ \bottomrule
\end{tabular}
\egroup
\tablefoot{\\
\tablefoottext{a}{The total integration time per half-wave plate (HWP) position is given by the multiplication of the detector integration time (DIT), the number of detector integrations (NDIT), and the number of NDIT summed over all dither positions (NINT).}\\
\tablefoottext{b}{Average airmass.}\\
\tablefoottext{c}{Average Differential Image Motion Monitor (DIMM) seeing in the optical with the standard deviation in parenthesis.}\\
\tablefoottext{d}{A neutral density (ND) filter with a $\sim$10$^{-1}$ transmissivity was used.}\\
}
\end{table*}

\subsection{SPHERE/ZIMPOL}\label{sec:zimpol_pdi}

HD~135344B was observed simultaneously with the \texttt{R\_PRIM}~filter ($\lambda_{\rm c}=\SI{0.626}{\micro\meter}$, $\Delta\lambda=\SI{0.149}{\micro\meter}$) and the \texttt{I\_PRIM}~filter ($\lambda_{\rm c}=\SI{0.790}{\micro\meter}$, $\Delta\lambda=\SI{0.153}{\micro\meter}$) in the \mbox{\texttt{SlowPol}} detector mode of \mbox{ZIMPOL}. The instrument converts the polarization of the incoming light into two intensity modulated beams with a polarization modulator and a polarization beam splitter. Each of the two beams is registered on an individual demodulating detector with a field of view of $3\ffarcs5 \times 3\ffarcs5$. In \mbox{\texttt{SlowPol}} mode, the de-modulation frequency of the two orthogonal polarization directions is reduced to $\sim$27~Hz which is a factor of $\sim$36 lower than in \mbox{\texttt{FastPol}} mode \citep{bazzon2012,schmid2012}. This enables a longer readout time for each individual frame and a factor of $\sim$10 reduction of readout noise. This mode is particularly useful for background limited regions of the image. The effective pixel scale is $3.5837 \pm 0.0006$ by $3.6045 \pm 0.0005$~mas and $3.5873 \pm 0.0005$ by $3.6081 \pm 0.0006$ mas for camera~1 ($I$-band) and camera~2 ($R$-band), respectively (Ginski et al., in prep.). The difference in pixel scale is caused by the toric mirrors in the common path infrastructure \citep[CPI;][]{hugot2012} that introduce anamorphic magnification differences between the horizontal and vertical detector direction.

The \mbox{ZIMPOL} observations were carried out with a detector integration time (DIT) of 10~s to reach a high signal-to-noise (S/N) but without strongly saturating the detector. No coronagraph was used in order to be sensitive to the innermost disk regions but the data is not usable inside approximately 30~mas due to saturation effects. The central octopole pattern that is visible between 30 and 100~mas in both the $Q_\phi$ and $U_\phi$ images (see Fig.~\ref{fig:images_zimpol}) is likely an instrumental effect and/or a low-wind effect which reduces the S/N very close to the star. Unsaturated frames were obtained with the use of an additional neutral density filter ($\sim$10$^{-1}$ transmissivity). This allows us to estimate the angular resolution of the images and to do a photometric calibration (see Appendix~\ref{sec:photometric_calibration}). The average Differential Image Motion Monitor (DIMM) seeing during the observations was $\sim$0\ffarcs7.

Four half-wave plate (HWP) positions were cycled ($0\degr$, $45\degr$, $22.5\degr$, and $67.5\degr$) to enable construction of Stokes \mbox{+Q}, \mbox{-Q}, \mbox{+U}, and \mbox{-U}, respectively. Each HWP orientation was successively used for eight integrations. The observations comprised three dithering positions with steps of 71~mas to minimize the contribution of bad pixels. A total number of 26 complete polarimetric cycles were obtained and a single polarimetric cycle and single dithering position was used for the total intensity measurement. The total integration time per HWP position was 960~s and an additional 40~s per HWP position with the neutral density filter (see Table~\ref{table:summary_observations}). 

The raw frames of the \mbox{ZIMPOL} observations were split into the two beams, $R$- and $I$-band, and the polarization states were extracted from the odd and even detector rows \citep{thalmann2008}. Standard data reduction routines were applied including flat field, bias frame, and bad pixel correction \citep{avenhaus2014b}. The stellar position was determined using an asymmetric 2D Gaussian fit because of the non-square nature of the \mbox{ZIMPOL} pixels. We have used a correction of $-1.188\degr$ for the rotational offset of the HWP with respect to true north (see Eq.~\ref{eq:observation_pa}) which is determined by minimizing the integrated signal in the $U_\phi$ image. The frames have been upscaled by a factor of two in vertical direction to make the pixels square, they were centered with the derived stellar location, and non-linear pixels have been masked out. Instrumental polarization is corrected by equalizing the integrated flux of the ordinary and extraordinary beam between 0\ffarcs1 and 1\ffarcs0 \citep{avenhaus2014b}. $Q_\phi$ and $U_\phi$ frames (see Sect.~\ref{sec:stokes_parameters}) were obtained for each HWP cycle and combined with a mean-stacking. Astrometric calibration was performed on the final $Q_\phi$ and $U_\phi$ images (Ginski et al., in prep.). A more exhaustive description of the data reduction technique is provided in the appendix of \citet{avenhaus2014b}.

The angular resolution of the \mbox{ZIMPOL} observations is determined from a reduced and non-saturated total intensity frame. The full width half maximum (FWHM) of the point spread function (PSF) is 33.5~mas and 28.5~mas for $R$- and $I$-band, respectively, which are comparable due to the smaller diffraction limit in $R$-band but a higher Strehl ratio in $I$-band.

\subsection{SPHERE/IRDIS}\label{sec:irdis_dpi}

In addition to the \mbox{ZIMPOL} PDI observations, we carried out \mbox{IRDIS} dual-polarization imaging (DPI) observations at two different epochs approximately one month apart. We note that PDI and DPI are equivalent techniques but there is an instrumental difference between \mbox{ZIMPOL} and \mbox{IRDIS} \citep{thalmann2008,langlois2014}. We observed HD~135344B with the \texttt{BB\_Y}~filter ($\lambda_{\rm c}=\SI{1.043}{\micro\meter}$, $\Delta\lambda=\SI{0.140}{\micro\meter}$) and without coronagraph during a HWP experiment which resulted in an on-source integration time of 151~s per HWP position with a DIT of 0.84~s. The \mbox{IRDIS} detector has a pixel scale of approximately 12.26~mas per pixel and a $11\arcsec \times 11\arcsec$ field of view. The HWP that controls the orientation of the polarization was cycled through $0\degr$, $45\degr$, $22.5\degr$, and $67.5\degr$ to obtain the linear components of the Stokes vector. We completed seven polarimetric cycles with an average DIMM seeing of $\sim$0\ffarcs9 (see Table~\ref{table:summary_observations}).

A deeper \mbox{IRDIS} observation was done with the \texttt{BB\_J}~filter ($\lambda_{\rm c}=\SI{1.245}{\micro\meter}$, $\Delta\lambda=\SI{0.240}{\micro\meter}$) for which we used a small apodized Lyot coronagraph with an inner working angle (IWA) of 80~mas. The total integration time per HWP position was 1152~s with a DIT of 32~s corresponding to 12 completed polarimetric cycles. A non-coronagraphic (unsaturated) frame was obtained at the start and the end of the observations with a 0.84~s DIT and an additional neutral density filter with $\sim$10$^{-1}$ transmissivity. Seeing conditions were good ($\sim$0\ffarcs7) and the spiral arms were visible after a single detector integration.

The ordinary and extraordinary beam were extracted from the raw frames of the \mbox{IRDIS} detector after which a dark frame, flat field, and bad pixel correction were applied \citep{avenhaus2014b}. To accurately determine the position of the star behind the coronagraph we used dedicated center calibration frames at the beginning and the end of the science sequence. These are coronagraphic images in which a periodic 2D modulation is applied to the
deformable mirror that causes four equidistant echoes of the stellar PSF to appear in a cross-shaped pattern outside of the coronagraph \citep{marois2006,sivaramakrishnan2006}. A linear interpolation between the two center frames is used to determine the stellar position in all frames. For the non-coronagraphic $Y$-band observations, we have determined the stellar position by fitting a 2D Moffat function to the saturated PSF profile. We have used a correction of $-1.7\degr$ for the rotational offset of the HWP with respect to true north (see Eq.~\ref{eq:observation_pa}).

Similar to the \mbox{ZIMPOL} observations, the instrumental polarization of the $J$-band observation is corrected by equalizing the ordinary and extraordinary beam. The instrumental polarization of the $Y$-band observations is determined by integrating the flux in an annulus with an inner radius of 6 pixels and outer radius of 9 pixels centered on the star for Stokes~$Q$ and $U$ separately. The corresponding total intensity frames are multiplied with the integrated flux value after which it is subtracted from the Stokes~$Q$ or $U$ image \citep{canovas2011}. The $Q_\phi$ and $U_\phi$ images were obtained from the Stokes~$Q$ and $U$ images with a mean stacking. The angular resolution of the \mbox{IRDIS} $J$-band image is 38.5~mas. No unsaturated frames were obtained during the $Y$-band observations and a determination of the angular resolution from the science data is not possible.

\subsection{Azimuthal Stokes parameters}\label{sec:stokes_parameters}

We convert the linear Stokes parameters, $Q$ and $U$, into their azimuthal counterparts, $Q_\phi$ and $U_\phi$, which are defined as \citep[cf.][]{schmid2006}:
\begin{equation}\label{eq:stokes_phi}
\begin{aligned}
Q_\phi &= Q\cos{2\phi}+U\sin{2\phi}\\
U_\phi &= Q\sin{2\phi}-U\cos{2\phi},
\end{aligned}
\end{equation}
where $\phi$ is the position angle of a location $(x,y)$ in the disk image with respect to the stellar position $(x_0,y_0)$. The position angle is given by
\begin{equation}\label{eq:observation_pa}
\phi = \arctan{\frac{y-y_0}{x-x_0}} + \theta,
\end{equation}
where $\theta$ corrects for small instrumental offsets such as an angular misalignment of the HWP. In this way, $Q_\phi > 0$ for single scattered light from protoplanetary dust grains that introduce positive polarization at the observed scattering angles. This implies that the polarization vectors point in azimuthal direction, in contrast to $Q_\phi<0$ which corresponds to polarization in radial direction. The polarized intensity, $PI=\sqrt{Q^2+U^2}$, is similar to the $Q_\phi$ flux when multiple scattering can be neglected. In that case, the $U_\phi$ image does not contain any scattered light from the disk and can be used to estimate the noise level of the $Q_\phi$ image. In contrast to $PI$, $Q_\phi$ does not contain any bias from computing the squares of $Q$ and $U$ when noise is present in the image \citep{schmid2006}.

\section{Results}\label{sec:results}

\subsection{Polarized light imagery}\label{sec:polarized_light_imagery}

\begin{figure*}
\centering
\includegraphics[width=17.37cm]{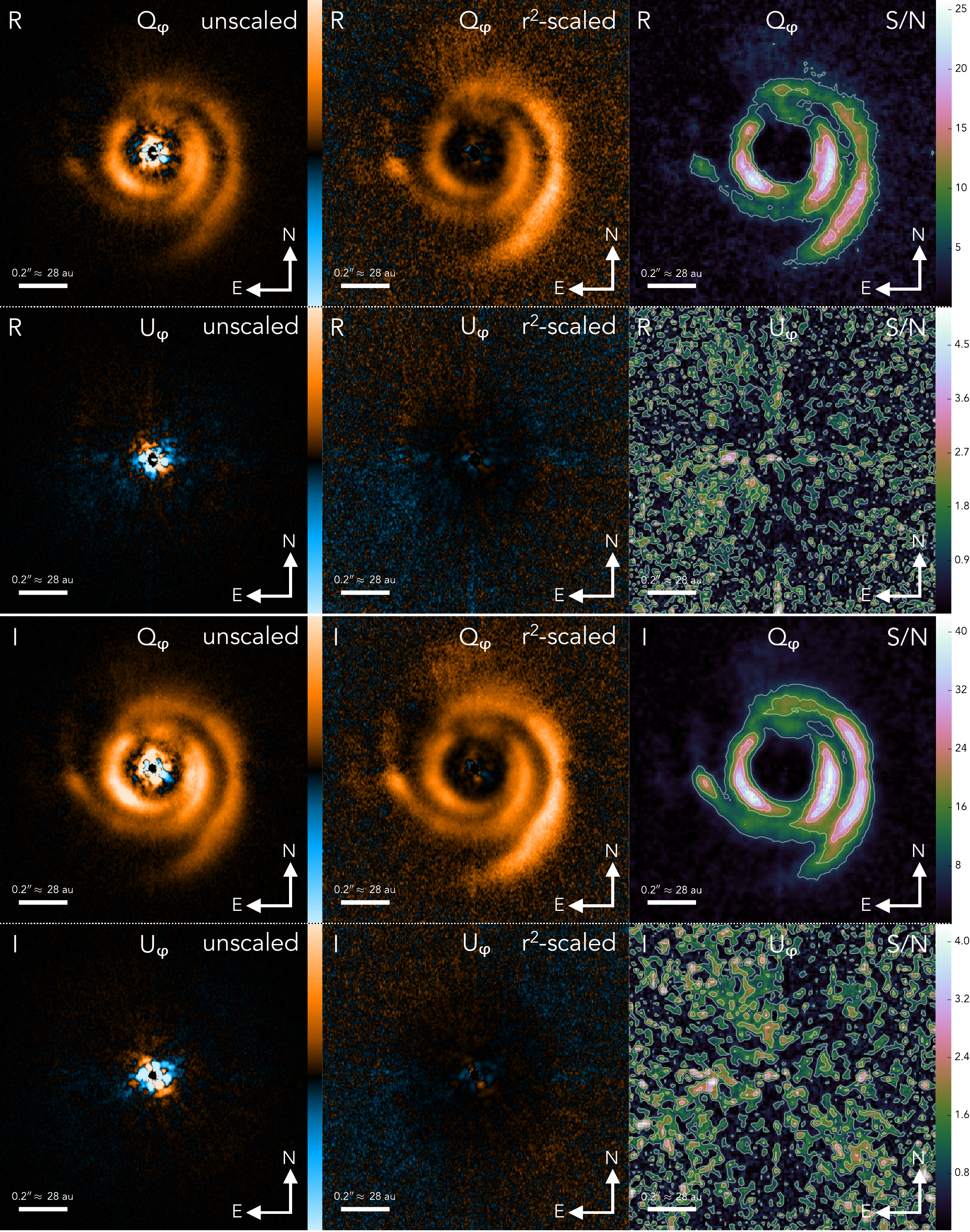}
\caption{\mbox{ZIMPOL} $R$-band (top two rows) and \mbox{ZIMPOL} $I$-band (bottom two rows) polarized intensity images. The field of view is $1\ffarcs4 \times 1\ffarcs4$ and the central star is positioned in the center of each image. The left column shows the $Q_\phi$ and $U_\phi$ polarized intensity images (see Sect.~\ref{sec:stokes_parameters}). The center column shows the $Q_\phi$ and $U_\phi$ images scaled with the deprojected distance squared from the star to each pixel for an inclination of $11\degr$ and a position angle of the major axis of $62\degr$. All images are shown on a linear color stretch, with companion $Q_\phi$ and $U_\phi$ images having the same minimum and maximum value. Orange corresponds to positive values, blue to negative values, and black is the zero point. The right column shows the signal-to-noise (S/N) maps of the $Q_\phi$ and $U_\phi$ images which are obtained from the $U_\phi$ images. The contour levels correspond to the standard deviation values that are shown on the right of the S/N map colorbar.}
\label{fig:images_zimpol}
\end{figure*}

\begin{figure*}
\centering
\includegraphics[width=\textwidth]{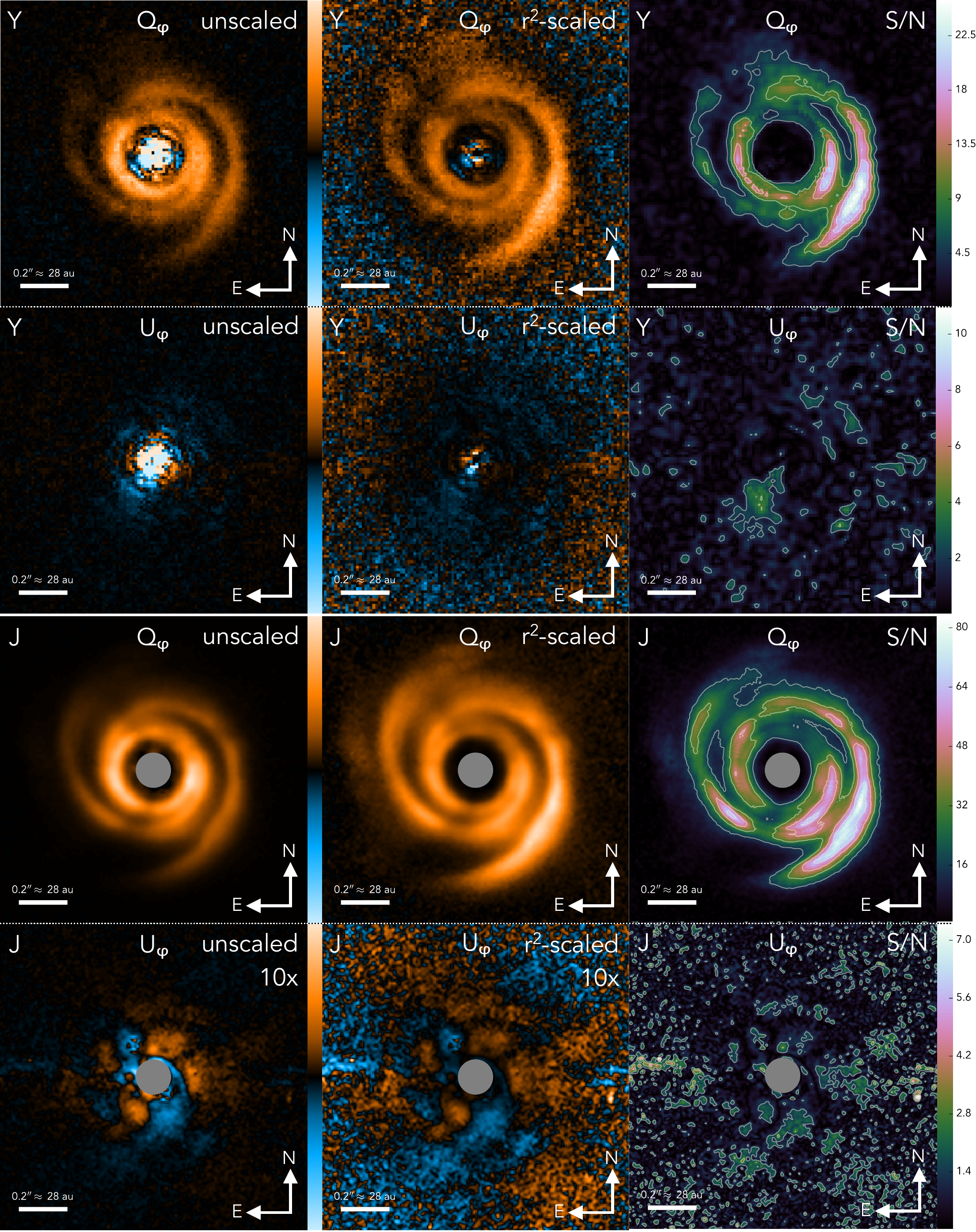}
\caption{\mbox{IRDIS} $Y$-band (top two rows) and \mbox{IRDIS} $J$-band (bottom two rows) polarized intensity images. The $J$-band $U_\phi$ images are shown with a ten times harder color stretch than the $J$-band $Q_\phi$ images. The 80~mas inner working angle of the coronagraph of the $J$-band observations has been masked out. See caption of Fig.~\ref{fig:images_zimpol} for further details.}
\label{fig:images_irdis}
\end{figure*}

Figures~\ref{fig:images_zimpol} and \ref{fig:images_irdis} show the \mbox{ZIMPOL} and \mbox{IRDIS} polarized intensity images, respectively, and the corresponding S/N maps of HD~135344B. The disk is clearly detected with all filters and a great amount of structure is seen with high S/N in the $Q_\phi$ images. The left columns show the unscaled $Q_\phi$ and $U_\phi$ polarized intensity images and the center columns the stellar irradiation corrected ($r^2$-scaled) $Q_\phi$ and $U_\phi$ images for which each pixel has been multiplied with the square of the deprojected distance to the stellar position at the center of each image. The correction compensates for the inverse-square law of the irradiation of the disk by the central star and attempts to provide an estimate of the spatial distribution of the dust grains in the disk surface. We have assumed a disk inclination of $11\degr$ and a position angle of the major axis of $62\degr$ (see Table~\ref{table:main_parameters}) and did not correct for the vertical extent of the disk.

Negative valued pixels are identified in the central regions of the unscaled $Q_\phi$ images, where PSF smearing effects \citep{avenhaus2014a} and low-wind effects dominate, and in the outer disk regions were the scattered light flux becomes comparable to the background noise level. The inner dust disk of HD~135344B is expected to be located at sub-au distance from the star \citep{carmona2014} and a direct detection is not possible. However, the signal pattern in the inner regions of the \mbox{ZIMPOL} images is expected to contain some of the smeared scattered light from the inner disk but at this point it is not clear how significant this is with respect to the instrumental signal.

The S/N maps in Figs.~\ref{fig:images_zimpol} and \ref{fig:images_irdis} are estimated from the $U_\phi$ images. We binned the pixel values in linearly spaced annuli (one pixel wide) around the star and calculated the standard deviation within each annulus which yields an estimate of the noise level in the $Q_\phi$ image as function of radial separation from the star. The S/N is calculated from the ratio of the absolute pixel values in the $Q_\phi$ image and the $U_\phi$ standard deviation in the corresponding annuli. For the $U_\phi$ S/N maps, we used the ratio of the absolute pixel values in $U_\phi$ image and the standard deviations from the $U_\phi$ annuli.

Multiple scattering in a moderate inclined and axisymmetric disk will produce non-azimuthal polarization in the $U_\phi$ image with values up to $\sim$5\% of the $Q_\phi$ flux \citep{canovas2015}. For a nearly face-on disk, the $U_\phi$ signal will be lower since forward scattering of large dust grains ($2\pi a \gtrsim \lambda$, with $a$ the grain radius and $\lambda$ the photon wavelength) reduces the number of multiple scattered photons that are directed towards the observer. A hard color stretch of the $J$-band $U_\phi$ image (see Fig.~\ref{fig:images_irdis}) is required to reveal the non-zero residual.

\begin{figure}
\centering
\resizebox{\hsize}{!}{\includegraphics{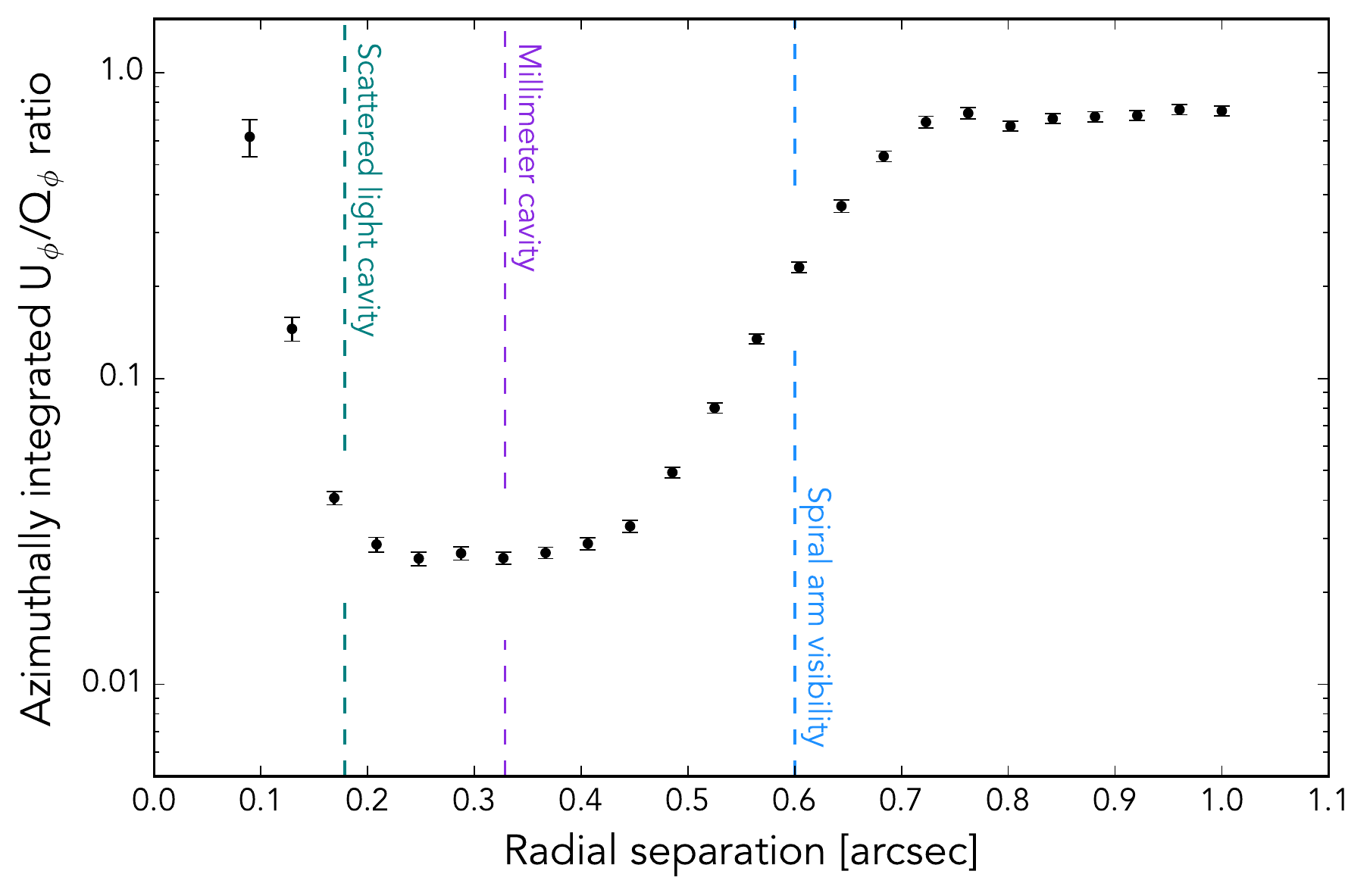}}
\caption{Azimuthally integrated ratio of the $U_\phi$ and $Q_\phi$ polarized surface brightness. It reaches a minimum of the order of a few percent around high S/N disk locations. The error bars show 5$\sigma$ as derived from the $U_\phi$ image. The dashed lines show the 25~au scattered light cavity, the 46~au millimeter cavity from \citet{andrews2011}, and the outer radius at which the spiral arms are visible.}
\label{fig:Qphi_Uphi}
\end{figure}

We determine the ratio of the azimuthally integrated $U_\phi$ and $Q_\phi$ signal in 25 linearly spaced annuli around the star between 0\ffarcs05 and 1\ffarcs0 (see Fig.~\ref{fig:Qphi_Uphi}). The number of bins is chosen such that the width of each annulus is approximately equal the angular resolution of the image. The $J$-band observation is used for this analysis because of the high S/N compared to the other filters. We have used the absolute pixel values because the annulus integrated $U_\phi$ signal from a face-on and axisymmetric disk will be zero. As expected, the ratio approaches unity both in the cavity ($\lesssim$0\ffarcs2) and the outermost disk region ($\gtrsim$0\ffarcs7) because the scattered light flux in the $Q_\phi$ image is similar to the noise level. The integrated $U_\phi$ signal is between 2.5\% and 3\% of the integrated $Q_\phi$ signal where the S/N is largest, i.e., where the $U_\phi/Q_\phi$ ratio reaches a minimum in Fig.~\ref{fig:Qphi_Uphi}. This might suggest that the effect of multiple scattering in the disk is non-negligible.

\subsection{Disk morphology and brightness asymmetries}\label{sec:disk_morphology}

\begin{figure}
\centering
\resizebox{\hsize}{!}{\includegraphics{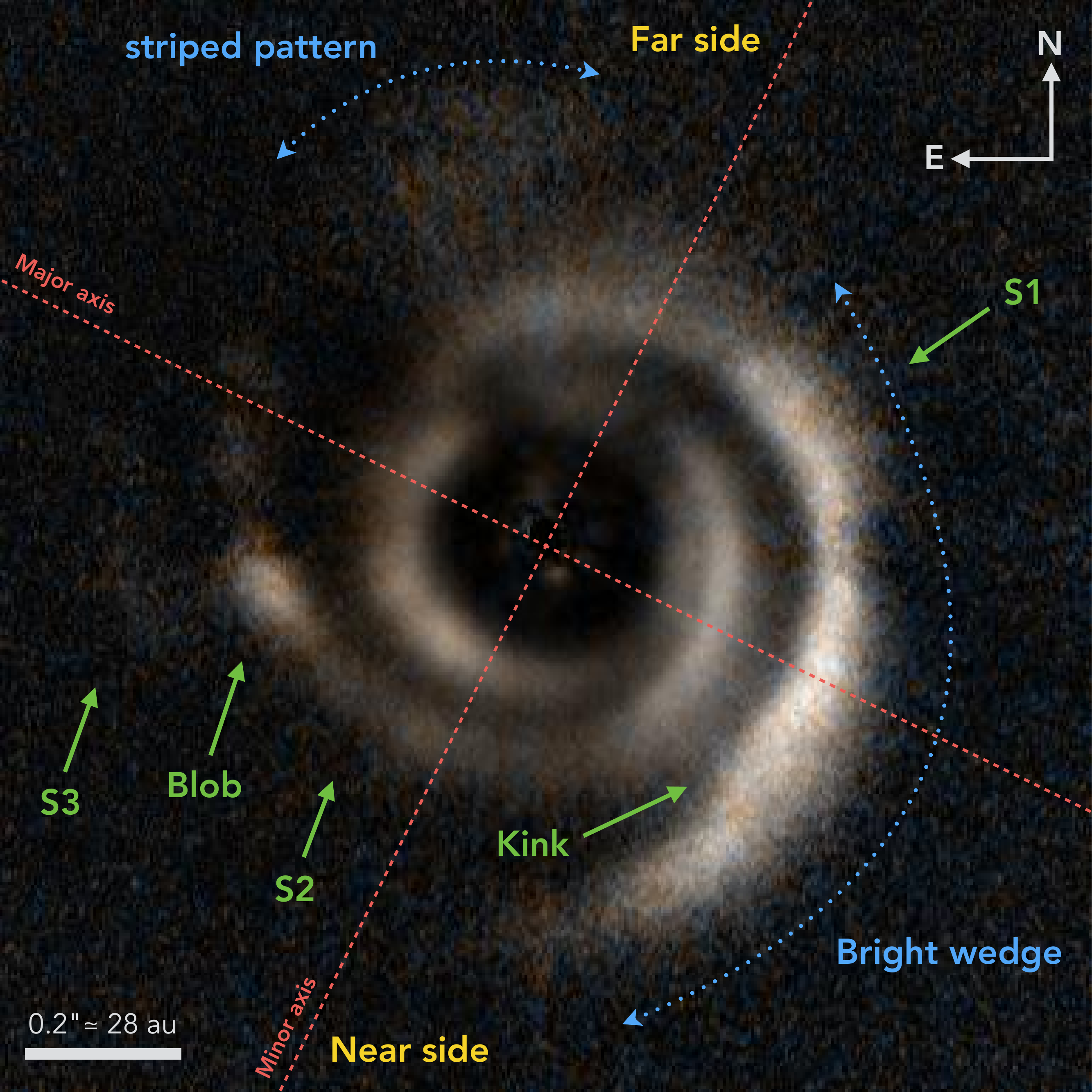}}
\caption{Color composite RGB image of \mbox{ZIMPOL} $R$-band (blue), \mbox{ZIMPOL} $I$-band (red), and the average of $R$- and $I$-band (green). The field of view is $1\ffarcs4 \times 1\ffarcs4$ and the main features that have been identified are labeled.}
\label{fig:zimpol_color}
\end{figure}

The disk is clearly detected from approximately 0\ffarcs1 up to 0\ffarcs7 in the $r^2$-scaled $Q_\phi$ images of both \mbox{ZIMPOL} and \mbox{IRDIS} (see Figs.~\ref{fig:images_zimpol} and \ref{fig:images_irdis}). Additionally, there is a tentative detection of the disk in $J$-band from 0\ffarcs7 to 1\ffarcs0 (see Sect.~\ref{sec:surface_brightness}). We have identified a number of structures in the $r^2$-scaled $Q_\phi$ images which are related to the morphology and surface brightness of the disk. The features have been labeled in Fig.~\ref{fig:zimpol_color} and will be described in more detail below. The figure shows a color composite image of the $r^2$-scaled ZIMPOL $R$- and $I$-band images. We have scaled the $I$-band image by a factor of 1.05 to correct for the larger photometric flux in $I$-band with respect to $R$-band \citep{coulson1995}.

Two prominent spiral arms, S1 and S2 \citep{muto2012,garufi2013}, are visible in all $Q_\phi$ images (see Fig.~\ref{fig:zimpol_color}) and can be traced from the outer disk cavity edge at $\sim$0\ffarcs2 to $\sim$0\ffarcs6 ($\sim$28-64~au). Spiral arm S1 separates from the cavity edge at a position angle (east of north) of $\sim$$80\degr$ and spiral arm S2 at a position angle of $\sim$$260\degr$, both spiral arms span an azimuthal range of $\sim$$240\degr$. The spiral arms are approximately azimuthally symmetric in their location and pitch angle and have a surface brightness contrast of a factor of 2-3 with the background disk in the $r^2$-scaled images. There appears to be a spiral arm branch on the east side of the disk, also detected by \citet{garufi2013}, which we label S3 (see also the $r^2$-scaled $J$-band image in Fig.~\ref{fig:images_irdis} where this feature is better visible). A kink in the S1 spiral arm is visible at a position angle of $225\degr$ (see Fig.~\ref{fig:zimpol_color}) which is identified by the sudden change in pitch angle of the spiral arm. We will discuss this feature in Sect.~\ref{sec:origin_spiral_arms}. We infer that the near side is along the southern minor axis because of the preceding and receding CO lines \citep{perez2014} and assuming that the observed spiral arms are trailing.

The scattered light cavity, initially revealed by \citet{garufi2013}, is clearly visible and has an approximate radius of 0\ffarcs18 (25~au). The cavity edge of the outer disk will be studied in the context of spatial segregation of dust grain sizes in Appendix~\ref{sec:outer_disk_cavity_edge}. Multiple surface brightness depressions, which are unlikely to be related to changes in disk or dust properties, are in particular well detected in the high S/N $J$-band observation. These features are likely shadows that are cast by the innermost disk regions. We will identify the shadow features and elaborate on their detection in Sect.~\ref{sec:shadows}.

All filters show in the $r^2$-scaled $Q_\phi$ image a positive surface brightness gradient in southwest direction. In particular the \mbox{ZIMPOL} images show a distinct bright wedge-shaped region (see Fig.~\ref{fig:zimpol_color}). The mean $r^2$-scaled pixel value in the bright wedge between 0\ffarcs15 and 0\ffarcs7 is a factor of 3.7 larger than the mean value from the same sized wedge in opposite direction. A possible interpretation of this brightness gradient will be discussed in Sect.~\ref{sec:disk_midplane}.

The \mbox{ZIMPOL} and \mbox{IRDIS} observations reveal a similar disk morphology yet several differences in surface brightness are visible. The flux contrast of the S2 spiral arm with the S1 spiral arm, measured on opposite sides of the major axis, is approximately a factor of five larger in the \mbox{ZIMPOL} $R$-band image compared to the \mbox{IRDIS} $J$-band image. The faintness of the S2 spiral arm in the \mbox{ZIMPOL} images could be a shadowing effect by the outer disk rim which will become optically thin higher above the midplane at shorter wavelengths. The \mbox{ZIMPOL} images show a bright blob in the S2 spiral arm at a position angle of $\sim$$100\degr$ which does not seem to be affected by the shadowing of the outer disk rim possibly due to a local perturbation of the surface density or temperature which may result in an increased scattered light flux. A striped pattern in azimuthal direction is present in the \mbox{ZIMPOL} data which is likely an artifact of the rotation of the telescope spiders during the observation (see Fig.~\ref{fig:zimpol_color}).

\subsection{Disk color in polarized light}\label{sec:disk_color}

The fraction of stellar light that scatters from the disk surface towards the observer depends on disk properties such as the pressure scale height and surface density, as well as dust properties such as the single scattering albedo, phase function, and polarizability (which are related to grain size, structure, and composition). We determine the disk scattering efficiency from the $R$-, $I$-, and $J$-band observations by calculating the disk integrated polarized intensity in the $Q_\phi$ image with an annulus aperture (0\ffarcs1--2\ffarcs0) centered on the star (only disk signal because the stellar halo is unpolarized) and we use a 2\ffarcs0 circular aperture to calculate the integrated total intensity of an (unsaturated) total intensity image (mainly stellar light). A correction is applied for the difference in DIT of the $Q_\phi$ polarized intensity frame and the total intensity frame. Additionally, we correct the total intensity frame for the transmissivity of the neutral density filter. We have neglected foreground extinction since it will affect the scattered light flux and stellar flux equally. For this analysis we could not include the $Y$-band observations since no unsaturated frames were obtained during the HWP experiment.

\begin{figure}
\centering
\resizebox{\hsize}{!}{\includegraphics{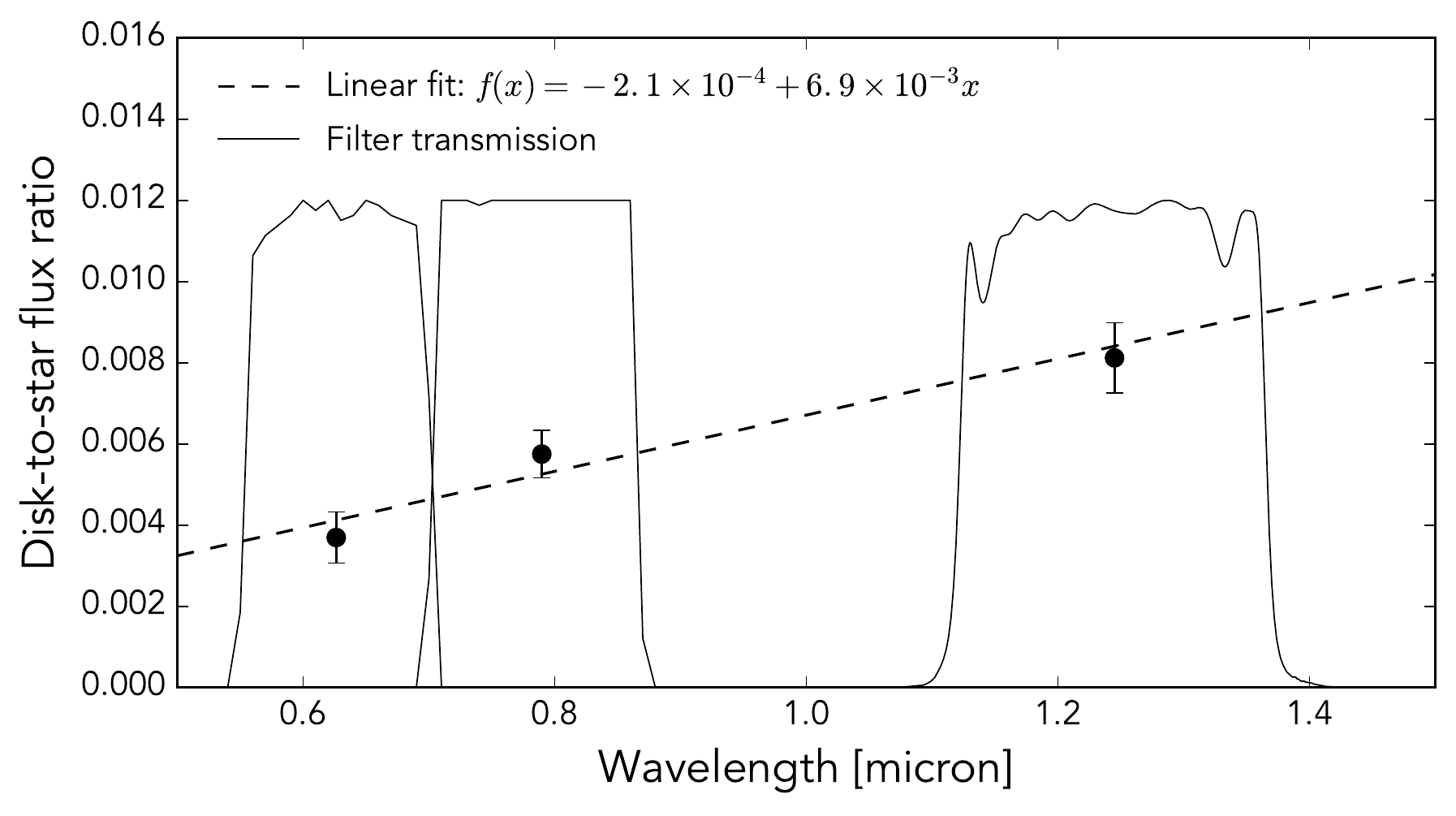}}
\caption{Disk-to-star flux ratio as determined from the $Q_\phi$ image, which only includes polarized scattered light from the disk, and an unsaturated total intensity image, which is dominated by the stellar halo. The three data points are the flux ratios in $R$-, $I$-, and $J$-band with 3$\sigma$ error bars. The dashed line shows a weighted least squares fit of a linear function and the solid lines are the normalized transmission curves of the three filters.}
\label{fig:disk_star_flux}
\end{figure}

Figure~\ref{fig:disk_star_flux} shows the disk-to-star flux ratio of the $R$-, $I$-, and $J$-band images. The uncertainties are derived from the $U_\phi$ image by propagating the error on each pixel value (see S/N maps of Figs.~\ref{fig:images_zimpol} and \ref{fig:images_irdis}) to an error bar on the integrated flux ratio. A weighted least squares fit of a linear function shows that the disk scattering efficiency increases towards longer wavelengths. If we assume that the degree of polarization is approximately the same for the three filers (since it is only weakly dependents on wavelength), then we can conclude that the disk integrated polarized intensity is intrinsically red in color consistently in $R$-, $I$-, and $J$-band (see also the red color of the composite \mbox{ZIMPOL} image in Fig.~\ref{fig:zimpol_color}). We note that the scattering geometry in the disk is very similar in the optical and near-infrared although near-infrared observations probe slightly deeper in the disk surface.

For sub-micron sized dust grains, the scattering cross section will be significantly larger at shorter wavelengths due to the $\lambda^{-4}$ wavelength dependence in the Rayleigh regime. We observe an opposite effect which indicates that larger dust grains ($2\pi a\gtrsim\lambda$) dominate the scattering opacity in the disk surface. Large dust grains give the disk a red appearance in scattered light because the shorter wavelength photons scatter more strongly in forward direction (i.e., into the disk) which makes the disk more faint at shorter wavelengths \citep{mulders2013a}. Large dust grains in the disk surface need to have an aggregate structure which prevents them from settling efficiently towards the disk midplane. A detailed multi-wavelength study of the optical properties of compact dust aggregates by \citet{min2016} showed that the effective albedo (scattering albedo with the forward scattering peak of the phase function excluded) of micron-sized aggregate dust grains increases from optical to near-infrared wavelengths.

\subsection{Surface brightness profiles}\label{sec:surface_brightness}

\begin{figure*}
\centering
\includegraphics[width=\textwidth]{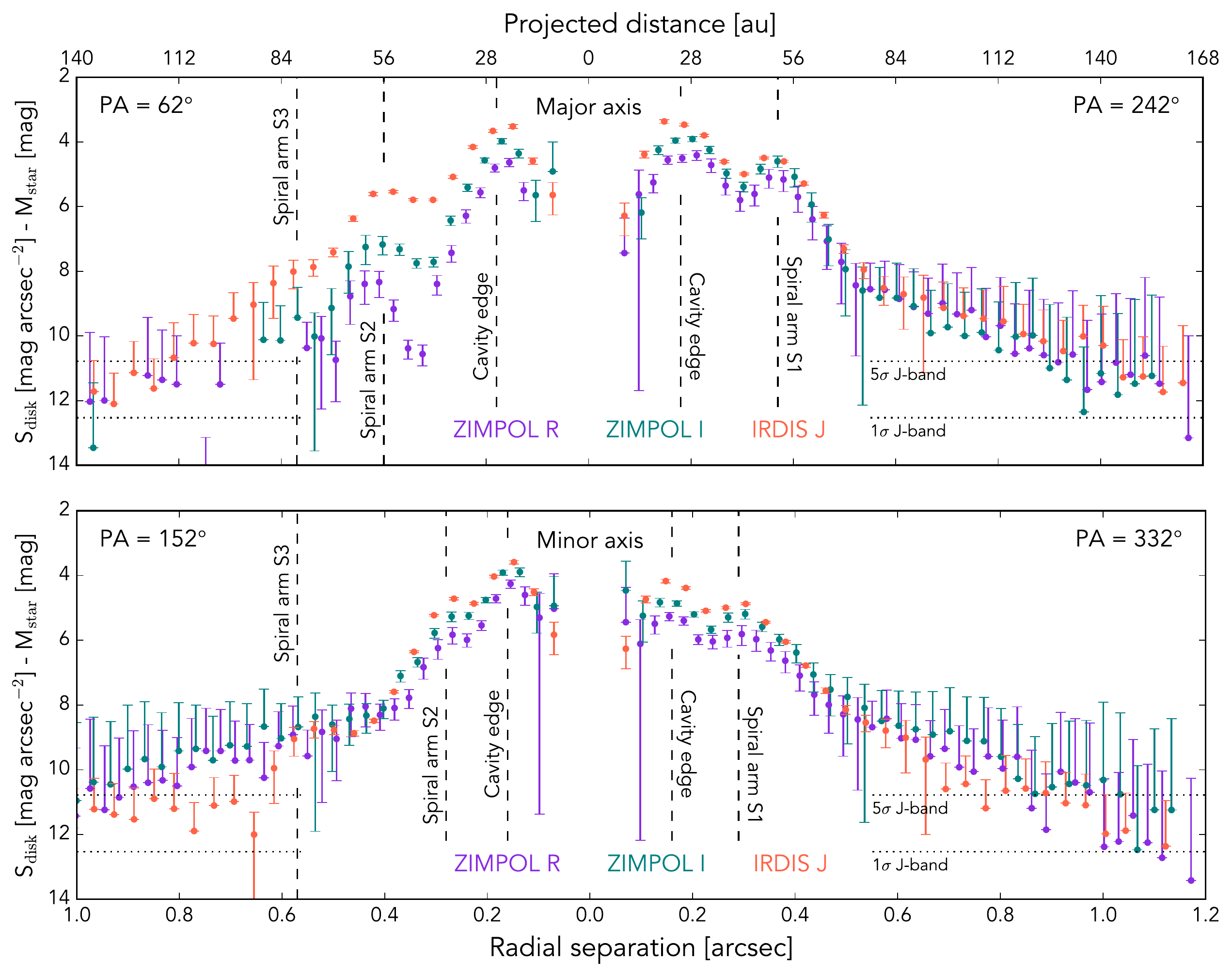}
\caption{Polarized surface brightness profiles along the major axis (top) and minor axis (bottom) of the HD~135344B disk. The position angle of the axis (measured east of north) is shown in the top left/right and the colors indicate different filters. Data points show the mean $Q_\phi$ surface brightness and the error bars show 1$\sigma$ uncertainties determined from the corresponding $U_\phi$ image. We have rejected bins in which the mean $Q_\phi$ count rate is negative as well as lower limit error bars that correspond to negative flux values. We note that the error bars do not include the uncertainty from the photometric calibration. The dotted lines show 1$\sigma$ and 5$\sigma$ background levels in the $Q_\phi$ $J$-band image at large separation from the star.}
\label{fig:surface_brightness_profile}
\end{figure*}

We use the (unscaled) $Q_\phi$ images to obtain polarized surface brightness profiles along the major and minor axis of the disk which are shown in Fig.~\ref{fig:surface_brightness_profile}. A radial cut is made through the disk with an azimuthal width of $10\degr$ and 41, 35, and 30 linearly spaced bins in radial direction between 0\ffarcs05 and 1\ffarcs0 for $R$-, \mbox{$I$-,} and $J$-band, respectively. The number of bins has been chosen such that the bin width is approximately equal to the angular resolution of the image. The error bars show 1$\sigma$ deviation in the corresponding bins of the unscaled $U_\phi$ images. The $Y$-band observation is excluded from the surface brightness profiles because no unsaturated total intensity frames were obtained which made a photometric calibration not possible.

Figure~\ref{fig:surface_brightness_profile} shows the difference of the disk polarized surface brightness and the stellar magnitude in the same filter (see Table~\ref{table:main_parameters}). In this way, each surface brightness profile is normalized to the stellar flux and depends mainly on the scattering properties of the dust grains \citep[e.g.,][]{quanz2012}. We note that the \mbox{IRDIS} photometric calibration has a larger uncertainty than the \mbox{ZIMPOL} photometric calibration because we derive a zero-point for the $J$-band observation from the \mbox{2MASS} stellar magnitude and the detector integrated total intensity with only applying a correction for the filter response (see Appendix~\ref{sec:photometric_calibration} for more details). Therefore, comparing the surface brightness between different wavelengths can only be done approximately. In general, the scattered light is red in color, consistent with the disk integrated color obtained in Sect.~\ref{sec:disk_color}, with local deviations from this overall trend. The dotted lines in Fig.~\ref{fig:surface_brightness_profile} show 1$\sigma$ and 5$\sigma$ of the background noise in the $J$-band $Q_\phi$ image which has been determined with a 100~pixel radius aperture at large separation from the star.

The surface brightness profiles are overall similar in shape and the disk is clearly detected up to $\sim$0\ffarcs7. Two evident maxima are visible along each direction with the innermost maximum at $\sim$0\ffarcs2 corresponding to the cavity edge of the outer disk, possibly intertwined with a spiral arm. The second surface brightness maximum is located around $\sim$0\ffarcs3--0\ffarcs4 which is caused by a spiral arm. The S2 spiral arm is relatively faint around the major axis in the \mbox{ZIMPOL} images which was also seen in Fig.~\ref{fig:zimpol_color}. The profiles along the $62\degr$ and $152\degr$ position angles show a third maximum around $\sim$0\ffarcs55 which was also detected by \citet{garufi2013}. It is most visible in the $J$-band surface brightness profile as a result of the high S/N disk detection. On the west side of the disk (${\rm PA}=242\degr$ and ${\rm PA}=332\degr$), there is a tentative detection of the disk between 0\ffarcs7 and 1\ffarcs0. In that region, the $J$-band flux is comparable to the 5$\sigma$ level of the background noise in $Q_\phi$.

The slope of the surface brightness profile beyond the spiral arms is determined by the shape of the unperturbed disk surface. Therefore, we fit the azimuthally averaged $Q_\phi$ profile between 0\ffarcs7 and 0\ffarcs9 with a power law function. An azimuthally averaged profile was used in order to enhance the faint scattered light flux at large disk radii. We performed a weighted least squares fit of a linear function in log-log space with the weights given by the standard deviations of the $U_\phi$ bins. The best-fit results of the power law exponents are $-4.33\pm0.11$, $-3.83\pm0.15$, $-1.22\pm0.40$, and $-2.71\pm0.07$ for $R$-, $I$-, $Y$-, and $J$-band, respectively. The values indicate that the surface height at which the disk in radial direction reaches an optical depth of unity might have a flat and not a flaring shape at disk radii $\gtrsim$100~au \citep{whitney1992}.

\subsection{Shadow features and indications for variability}\label{sec:shadows}

\begin{figure}
\centering
\resizebox{\hsize}{!}{\includegraphics{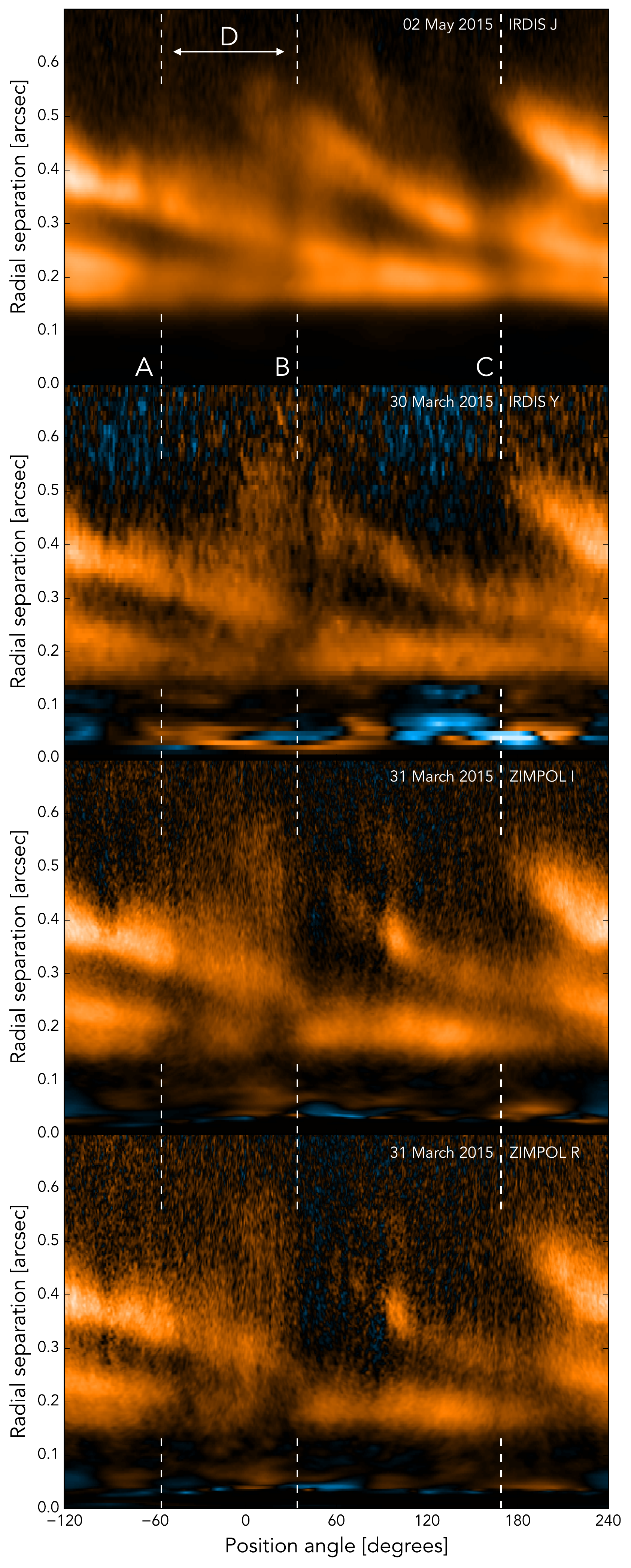}}
\caption{Polar projections of the $r^2$-scaled $Q_\phi$ images for $J$-, $Y$-, $I$-, and $R$-band (top to bottom). The vertically dashed lines correspond to shadow lanes A, B, and C and are identical to the dashed lines in Fig.~\ref{fig:model}. The identified shadow features have been labeled in the $J$-band image. North corresponds to ${\rm PA}=0\degr$ and positive position angles are measured east from north.}
\label{fig:polar_mapping}
\end{figure}

Figure~\ref{fig:polar_mapping} shows a projection of the $r^2$-scaled $Q_\phi$ images from Figs.~\ref{fig:images_zimpol} and \ref{fig:images_irdis} onto a polar coordinates grid. The polar projections point out more clearly surface brightness changes in radial and azimuthal direction and we identify multiple local depressions in surface brightness which appear over a large radial distance with a modest azimuthal gradient. It seems unlikely that this is related to changes in dust properties and/or scattering geometry given the near face-on orientation of the disk (inclination is $11\degr$), as well as the large radial extent and the azimuthal width of the local minima in surface brightness. Therefore, we interpret the surface brightness depressions as shadows that are cast by the innermost disk regions.

\begin{figure*}
\centering
\resizebox{\hsize}{!}{\includegraphics{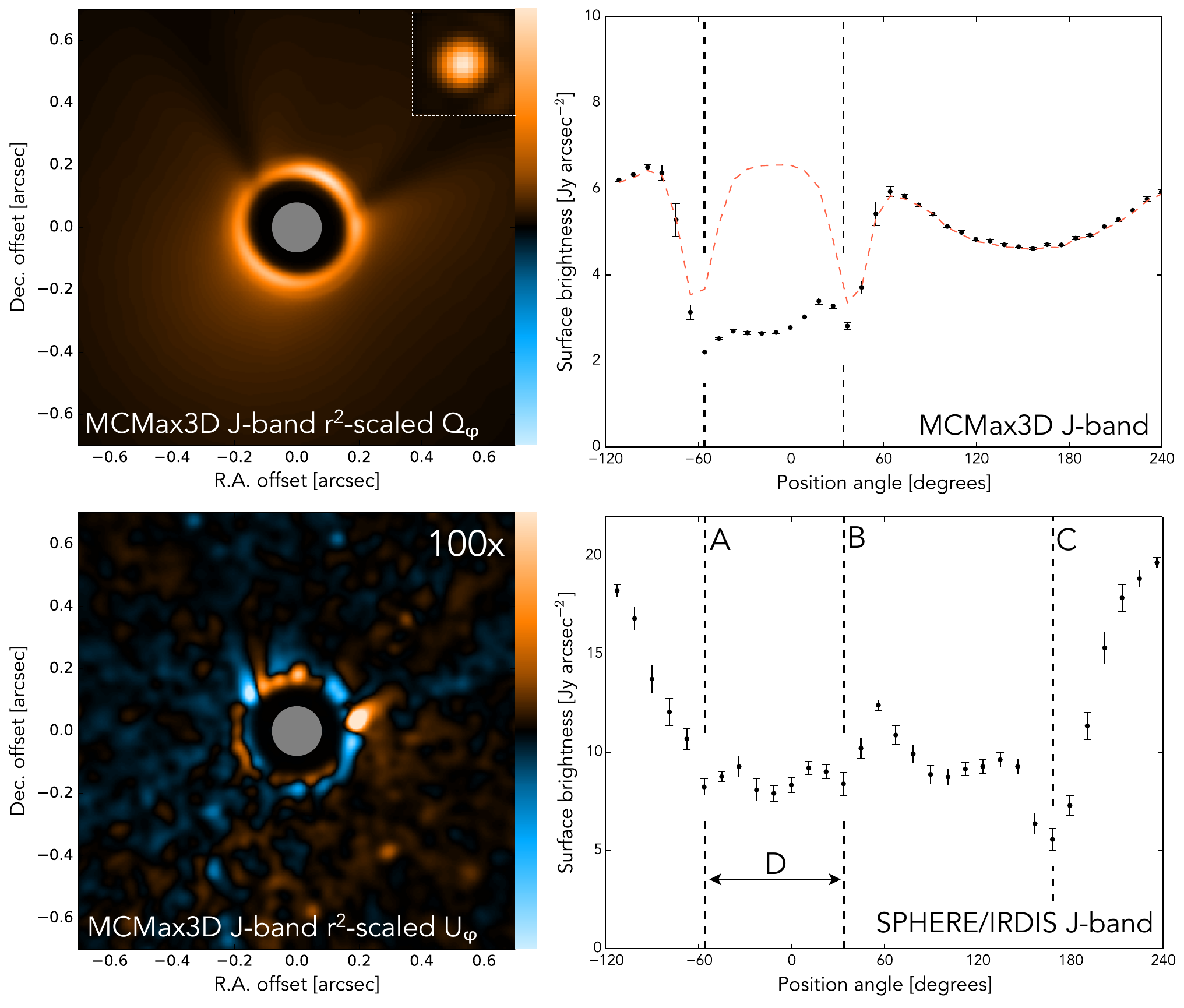}}
\caption{Synthetic $Q_\phi$ (top left) and $U_\phi$ (bottom left) polarized intensity images and the radially collapsed surface brightness from the radiative transfer model (top right) and the IRDIS $J$-band image (bottom right). The inset of the synthetic $Q_\phi$ image shows the PSF core of the $J$-band observation. The \mbox{MCMax3D} radiative transfer model has been constructed to qualitatively reproduce shadow features A, B, and D. Both images have been scaled with the squared distance from each pixel to the star and are shown on a linear color stretch with the $U_\phi$ color stretch 100~times harder than the $Q_\phi$ stretch. The 80~mas inner working angle of the coronagraph of the $J$-band observation has been masked out. The red dashed line in the top right figure shows the same radiative transfer model but without the warped disk region (zone 2). The vertically dashed lines show the locations of the surface brightness minima of shadow features A, B, and C. The error bars of the integrated surface brightness show 1$\sigma$ of the corresponding position angle bin in the $U_\phi$ image.}
\label{fig:model}
\end{figure*}

In particular the $r^2$-scaled $Q_\phi$ $J$-band image shows with high S/N the locations and shapes of the shadow features. In the $J$-band polar projection (top image of Fig.~\ref{fig:polar_mapping}), we identify three shadow lanes (A, B, and C) at position angles of approximately $-56\degr$, $34\degr$, and $169\degr$, respectively, which manifest themselves as dark vertical bands in the polar projection (dashed lines). Additionally, a broader shadow (D) is identified which is bound by shadow lanes A and B. The decrease in surface brightness at the shadow locations is quantified in the bottom right of Fig.~\ref{fig:model} which shows the radially integrated surface brightness of the unscaled $Q_\phi$ $J$-band image. The figure shows that the location of shadowed region D is bound by shadow lanes A and B but it is distinct from shadows A and B in terms of integrated surface brightness.

The \mbox{SPHERE} PDI observations were carried out during two different epochs which were approximately one month apart (see Table~\ref{table:summary_observations}). Here we refer to the $R$-, $I$-, and $Y$ band observations as epoch~1 and the $J$-band observations as epoch~2. Interestingly, a comparison of the polar projections of the two epochs in Fig.~\ref{fig:polar_mapping} shows that shadow feature C is present in the $J$-band image but appears to be absent in the image projections of epoch~1. A possible explanation for shadow features A, B, and D is a warped inner disk. We will explore this scenario with a radiative transfer model in Sect.~\ref{sec:radiative_transfer} and discuss the interpretation of the shadows in Sect.~\ref{sec:shadow_origin}.

\section{Modeling}\label{sec:modeling}

\subsection{Radiative transfer: shadows from a warped disk}\label{sec:radiative_transfer}

\begin{table}
\caption{\mbox{MCMax3D} model parameters}
\label{table:model_parameters}
\centering
\bgroup
\def\arraystretch{1.25}
\begin{tabular}{L{2.5cm} | C{1.5cm} C{1.5cm} C{1.5cm}}
\hline\hline
Parameter & Zone 1 & Zone 2 & Zone 3 \\
\hline
R$_{\rm in}$ [au] & 0.2 & 1 & 25 \\
R$_{\rm out}$ [au] & 1 & 25 & 200 \\
R$_{\rm tap}$\tablefootmark{a} [au] & 70 & 70 & 70 \\
R$_{\rm round}$ [au] & 0.3 & - & 27 \\
w\tablefootmark{b} & 0.3 & - & 0.3 \\
\hline
M$_{\rm dust}$ [M$_\odot$] & $2 \times 10^{-9}$ & $2\times 10^{-11}$ & $4 \times 10^{-4}$ \\
$\epsilon$\tablefootmark{c} & 1 & 3 & 1 \\
H$_0$/r$_0$\tablefootmark{d} & 0.01 & 0.01 & 0.07 \\
r$_0$\tablefootmark{e} [au] & 1 & 1 & 30 \\
$\psi$\tablefootmark{f} & 0 & 0 & 0.25 \\
$\alpha$\tablefootmark{g} & $10^{-3}$ & $10^{-1}$ & $10^{-3}$ \\
\hline
a$_{\rm min}$ [\SI{}{\micro\meter}] & 0.01 & 0.01 & 0.01 \\
a$_{\rm max}$ [\SI{}{\micro\meter}] & 1 & 1 & 1000 \\
a$_{\rm pow}$ & -3.5 & -3.5 & -3.5 \\
\hline
i [deg] & -11.5 & -11.5 : 11 & 11 \\
PA [deg] & 75 & 75 : 62 & 62 \\
\bottomrule
\end{tabular}
\egroup
\tablefoot{\\
\tablefoottext{a}{Tapering-off radius (see Eq.~\ref{eq:surface_density}).}\\
\tablefoottext{b}{Roundness of the disk rim (see Eq.~\ref{eq:surface_density}).}\\
\tablefoottext{c}{Surface density power law index (see Eq.~\ref{eq:surface_density}).}\\
\tablefoottext{d}{Reference aspect ratio.}\\
\tablefoottext{e}{Reference radius.}\\
\tablefoottext{f}{Flaring index (see Eq.~\ref{eq:vertical_gas}).}\\
\tablefoottext{g}{Turbulent mixing strength \citep{shakura1973}.}\\
}
\end{table}

\begin{figure}
\centering
\resizebox{\hsize}{!}{\includegraphics{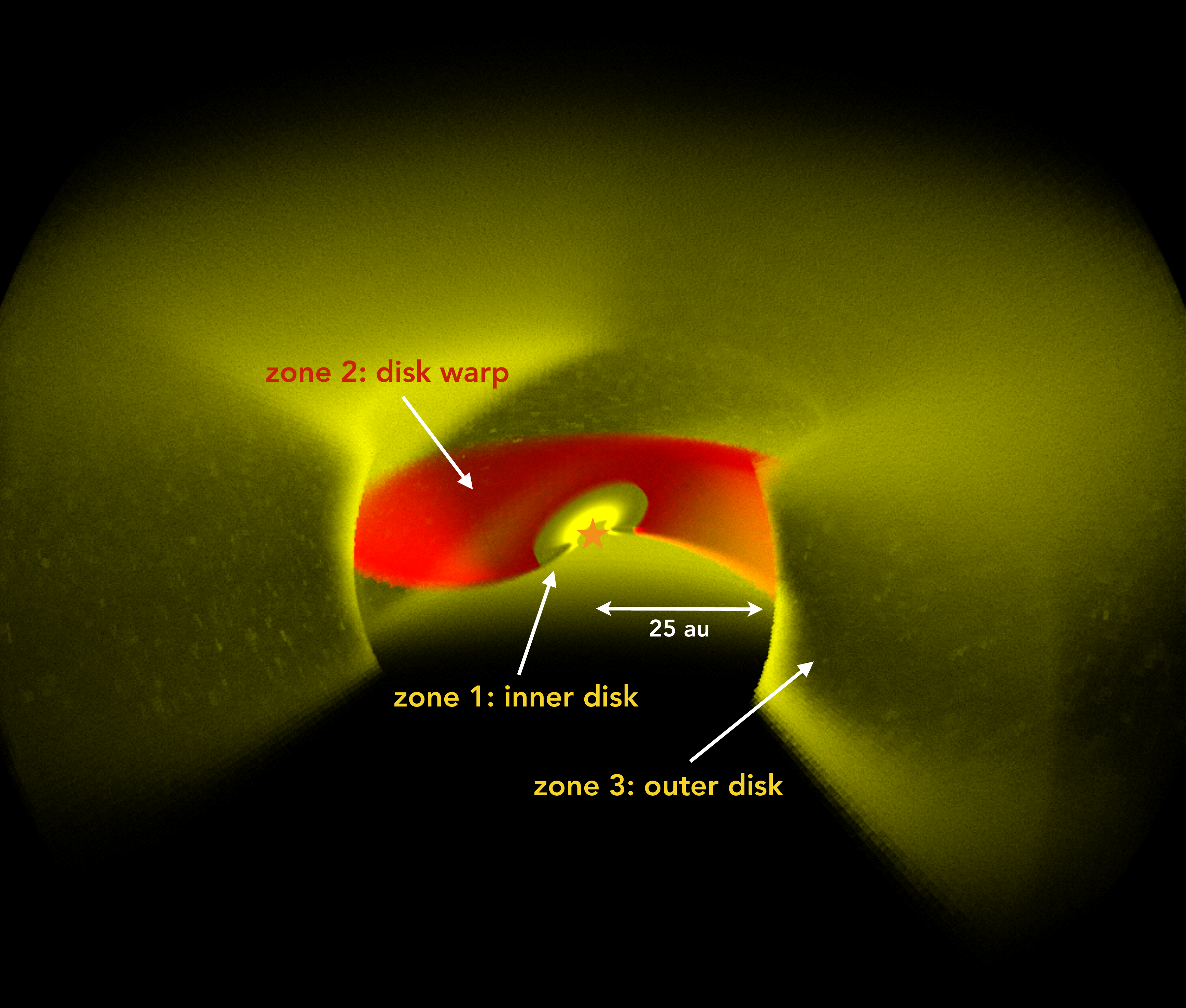}}
\caption{Architecture of the \mbox{MCMax3D} radiative transfer model (exaggerated for clarification of the disk warp). The misaligned inner disk (zone~1) and outer disk (zone~3) are shown in yellow. The warped disk region (zone~2), which transitions the orientation of the inner disk towards the outer disk, is shown in red.}
\label{fig:warp}
\end{figure}

A striking discovery of the \mbox{SPHERE} PDI observations is the presence of multiple shadows that are cast on the outer disk of HD~135344B. In Sect.~\ref{sec:shadows}, we identified four shadow features (A, B, C, and D) in the $J$-band $Q_\phi$ image (see Fig.~\ref{fig:polar_mapping}) and we showed that shadow feature C was only present in the second observation epoch. To explain shadow features A, B, and D, we propose a scenario in which the unresolved inner disk of HD~135344B is misaligned with respect to the outer disk and both are connected through a warped disk region.

To test the warped disk hypothesis, we use the 3D radiative transfer code \mbox{MCMax3D}, which fully includes multiple scattering and polarization, to qualitatively reproduce shadow features A, B, and D. Shadow feature C is not included in the model because it appears to be variable and its origin is uncertain (see the discussion in Sect.~\ref{sec:shadow_origin}). Also the spiral arms are not included since the shadow features are not affected by the presence of the spiral arms and including them would add unnecessary complexity.

We use the 3D version of the continuum radiative transfer code \mbox{MCMax} \citep{min2009} which calculates the thermal structure of the disk for a radially parametrized surface density distribution with a rounded disk rim \citep{hughes2008,mulders2013b}:
\begin{equation}\label{eq:surface_density}
\Sigma(r) = \Sigma_0 r^{-\epsilon} \exp{\left[ -\left(\frac{r}{R_{\rm tap}}\right)^{2-\epsilon} \right]} \exp{\left[ -\left(\frac{1-r/R_{\rm round}}{w}\right)^3 \right]},
\end{equation}
where $\Sigma_0$ is a normalization constant, $R_{\rm in} < r < R_{\rm out}$ the disk radius, $R_{\rm tap}$ the tapering-off radius, $\epsilon$ the surface density power law index, $R_{\rm round}$ the radius were the rounding of the surface density sets in, and $w$ a measure for how round the disk rim is. The surface density profile is scaled to the total dust mass, M$_{\rm dust}$, within each zone. The vertical density distribution is assumed to be Gaussian, in cylindrical coordinates given by
\begin{equation}\label{eq:vertical_gas}
\rho(r,z) = \frac{\Sigma(r)}{\sqrt{2}H(r)} \exp{\left[ -\frac{z^2}{2H(r)^2} \right]},
\end{equation}
where $H(r)/r = (H_0/r_0) (r/r_0)^\psi$ is the parametrization of the aspect ratio, $H(r)$ the pressure scale height, $H_0/r_0$ the aspect ratio at reference radius $r_0$, and $\psi$ the flaring index.

The grain size distribution of the dust is a power law with index $a_{\rm pow}$ and a minimum and maximum grain size of $a_{\rm min}$ and $a_{\rm max}$, respectively, for which we use values from Model~5 of \citet{carmona2014} (see Table~\ref{table:model_parameters}). The dust composition is 80\% silicates and 20\% carbon and the porosity of the grains is approximated with effective medium theory and set to 25\%. We consider the grains to be irregular in shape by setting the maximum volume void fraction used for the distribution of hollow spheres (DHS) to 0.8 \citep{min2005}. The composition, porosity, and irregularity are set to typical values from the European \mbox{FP7} project \mbox{DiscAnalysis} \citep[\mbox{DIANA};][]{woitke2016}. Grain size dependent dust settling is calculated with the prescription from \citet{woitke2016} who follow the method from \citet{dubrulle1995} but provide an adjustment for the parametrized vertical gas structure. The method assumes an equilibrium between gravitational settling and turbulent mixing and depends on a single free parameter, the turbulent mixing strength $\alpha$. The main parameters of the radiative transfer model are specified in Table~\ref{table:model_parameters}. We adopt the stellar temperature, luminosity, mass, and distance from Table~\ref{table:main_parameters}.

The radiative transfer model for HD~135344B consists of three different zones which are independent in their disk structure and dust properties. We refer to those as zone~1, 2, and 3 which correspond to the misaligned inner disk, the warped transition region, and the outer disk, respectively (see Fig.~\ref{fig:warp}). The location, shape, and surface brightness of shadow lanes A and B are mainly determined by the radial width, orientation, aspect ratio, and dust mass of the inner disk, as well as the aspect ratio and flaring of the outer disk. Shadow feature D is cast by the warped disk region and is mainly affected by the dust mass, surface density exponent, and the turbulent mixing strength of zone~2. Also the spatial scale on which the inner disk midplane transitions towards the outer disk midplane (see Eq.~\ref{eq:warp}) affects shadow feature~D.

The disk warp (zone~2) is modeled by transitioning the inner disk (zone~1) orientation stepwise towards the outer disk (zone~3) orientation. The inclination, $\theta_\mathrm{MCMax3D}$, of the warped part of the disk is given by
\begin{equation}\label{eq:warp}
\theta_\mathrm{MCMax3D} = \theta_\mathrm{in} + (\theta_\mathrm{out}-\theta_\mathrm{in}) \left[ \frac{R-R^\mathrm{out}_\mathrm{inner}}{R^\mathrm{in}_\mathrm{outer}-R^\mathrm{out}_\mathrm{inner}} \right]^{p_\mathrm{warp}},
\end{equation}
with $R^\mathrm{in,out}_\mathrm{inner}$ the inner/outer radius of the inner disk, $R^\mathrm{in,out}_\mathrm{outer}$ the inner/outer radius of the outer disk, and $p_\mathrm{warp}$ a parameter determining how quickly the inclination of the warped disk changes. We note that $\theta_\mathrm{MCMax3D}$ is the inclination in the \mbox{MCMax3D} model and not the inclination with respect to the detector plane. The azimuthal orientation in the warped part of the disk, $\phi_\mathrm{MCMax3D}$, is changing in a similar way from the inner disk towards the outer disk. A power law exponent of $p_\mathrm{warp}=0.2$ gave the best result for reproducing shadow feature~D.

Figure~\ref{fig:model} shows the synthetic $r^2$-scaled $Q_\phi$ and $U_\phi$ images in $J$-band as well as the radially integrated surface brightness of the synthetic and the observed $Q_\phi$ image. The synthetic Stokes $Q$ and $U$ images have been convolved with the PSF core from the $J$-band observation (see inset of the synthetic $Q_\phi$ image in Fig.~\ref{fig:model}) before converting them into their azimuthal counterparts. The radially integrated surface brightness is obtained from the unscaled $Q_\phi$ images by adding all pixels between 0\ffarcs2 and 0\ffarcs7 in 33 linearly spaced position angle bins. The number of bins is chosen such that the bin width at the inner radius approximately equals the angular resolution of the image. The error bars show 1$\sigma$ of the pixel values in the corresponding position angle bin of the $U_\phi$ image. The error bars of the high S/N \mbox{MCMax3D} image correspond to a real spread in the $U_\phi$ signal as a result of multiple scattering in the disk because the synthetic $Q_\phi$ and $U_\phi$ image are noise-free. For the observed $J$-band image, it is a combined effect of multiple scattering and noise residual (see Sect.~\ref{sec:polarized_light_imagery}).

The radiative transfer model reproduces qualitatively the location, shape, and the surface brightness contrast of shadow features A, B, and D both in the polarized intensity image and the radially collapsed surface brightness. The inner disk requires a $22\degr$ misalignment with the outer disk, a relatively small aspect ratio ($h_0 = 0.01$), and a dust mass of $2\times 10^{-9}$~M$_\odot$ to reproduce shadow features A and B. Reproducing shadow feature D requires a warped disk region (zone~2) with a steep surface density profile ($\epsilon=3$) and a dust mass of $2\times 10^{-11}$~M$_\odot$. This is consistent with the SED suggesting that the inner dust disk emission is optically thick in thermal emission but transitions to optically thin in the cavity. Furthermore, we used $\alpha=10^{-1}$ for zone~2 to make sure that the dust grains are vertically mixed. The effect of the warped disk region is shown in the radially integrated surface brightness of the \mbox{MCMax3D} model (top right of Fig.~\ref{fig:model}). The red dashed line shows the surface brightness of the same radiative transfer model but with zone~2 removed. In that case, only shadow features A and B are visible and the depression in surface brightness of shadow feature D is absent. Although shadow features A, B, and D are qualitatively reproduced, we note that the solution is not unique given the large number of free parameters that have an effect on the shadows. A more detailed study of each parameter is beyond the scope of this paper. Also, we did not attempt to fit the SED and the resolved millimeter emission.

\subsection{Spiral arms as tracers of protoplanets}\label{sec:spiral_arms}

Although several mechanism for the excitation of spiral arms exist (see the discussion in Sect.~\ref{sec:origin_spiral_arms}), in this section we will study in more detail the scenario in which the HD~135344B spiral arms have been excited by forming protoplanets. A low-mass protoplanet triggers density waves at the Lindblad resonances of the disk in which it is embedded \citep{ogilvie2002}. Interference of different azimuthal modes results in a one-armed spiral which can be approximated with linear perturbation theory when $M_{\rm p}/M_* << (H(r)/r)^3$, where $M_{\rm p}$ is the mass of the protoplanet, $M_*$ the stellar mass, $H(r)$ the pressure scale height, and $r$ the disk radius. Higher order perturbation terms become important when $M_{\rm p}/M_* \sim (H(r)/r)^3$ which can lead to the presence of secondary spiral arm \citep{juhasz2015,zhu2015}.

The shape of a planet-induced spiral arm as derived from density wave theory in the linear or weakly non-linear regime is given by \citep{rafikov2002}
\begin{equation}\label{eq:spiral}
\begin{aligned}
\phi(r) = & \phi_{\rm p} - \frac{{\rm sgn}(r-r_{\rm p})}{h_{\rm p}} \left(\frac{r}{r_{\rm p}}\right)^{1+\eta} \left[ \frac{1}{1+\eta}-\frac{1}{1-\zeta+\eta} \left( \frac{r}{r_{\rm p}} \right)^{-\zeta} \right] \\
& - \frac{{\rm sgn}(r-r_{\rm p})}{h_{\rm p}} \left( \frac{1}{1+\eta}-\frac{1}{1-\zeta+\eta} \right),
\end{aligned}
\end{equation}
where $\zeta$ is the power law index in the $\Omega \propto r^{-\zeta}$ disk rotation profile and $\eta$ the power law index that determines the steepness of the sound speed profile, $c_{\rm s} \propto r^{-\eta}$. The polar coordinates of the protoplanet location are $(r_{\rm p}, \phi_{\rm p})$ and the disk aspect ratio at the protoplanet location is given by $h_{\rm p}=c_{\rm s}(r_{\rm p})/(r_{\rm p}\Omega(r_{\rm p}))$. The pitch angle of the spiral arm depends mainly on the temperature at the launching point of the density wave.

\begin{table*}
\caption{Spiral arm best-fit results}
\label{table:spiral_best_fit}
\centering
\bgroup
\def\arraystretch{1.25}
\begin{tabular}{L{7cm} C{3.7cm} C{3.7cm}}
\hline\hline
Protoplanets inside the scattered light cavity & Prior & Best-fit value \\
\hline
Aspect ratio $h_{\rm S1}$ & [0.05, 0.4] & 0.4 \\
Planet S1 radial separation $r_{\rm S1}$ [au] & [0, 28] & 24.3 \\
Planet S1 position angle $\phi_{\rm S1}$ [deg] & [0, 360] & 10.4 \\
Planet S2 radial separation $r_{\rm S2}$ [au] & [0, 28] & 21.9 \\
Planet S2 position angle $\phi_{\rm S2}$ [deg] & [0, 360] & 245.2 \\
\hline
Protoplanets beyond the scattered light cavity & Prior & Best-fit value \\
\hline
Aspect ratio $h_{\rm S1}$ & [0.05, 0.4] & 0.16 \\
Planet S1 radial separation $r_{\rm S1}$ [au] & [28, 168] & 168.0 \\
Planet S1 position angle $\phi_{\rm S1}$ [deg] & [0, 360] & 52.2 \\
Planet S2 radial separation $r_{\rm S2}$ [au] & [28, 168] & 99.0 \\
Planet S2 position angle $\phi_{\rm S2}$ [deg] & [0, 360] & 354.8 \\
\hline
\bottomrule
\end{tabular}
\egroup
\end{table*}

\begin{figure}
\centering
\resizebox{\hsize}{!}{\includegraphics{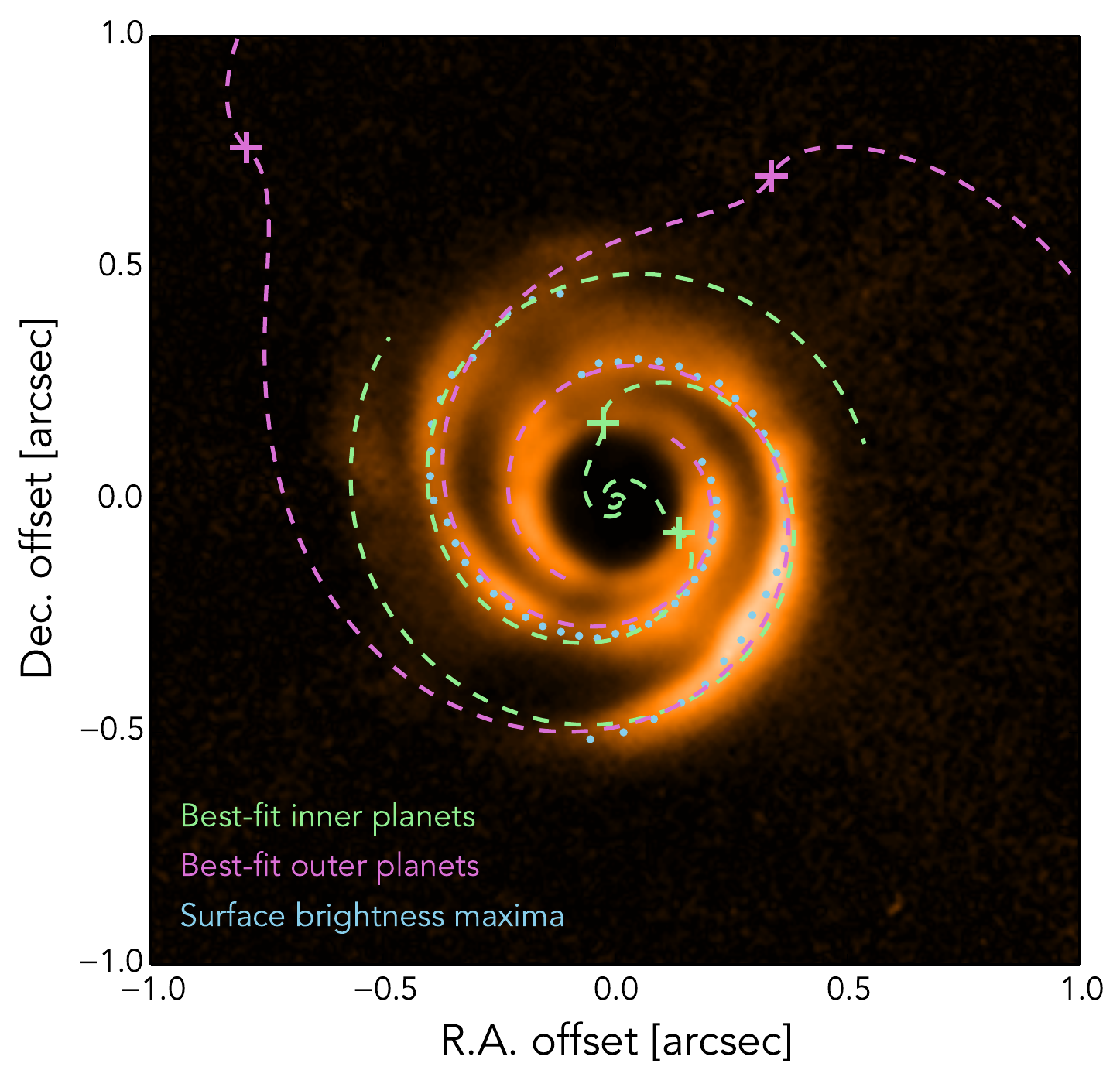}}
\caption{Best-fit spiral arm solutions for the protoplanets inside (green) and outside (purple) the scattered light cavity (see also Table~\ref{table:spiral_best_fit}). Blue points are the surface brightness maxima along each spiral arm that are used for the $\chi^2$ fitting. The plus symbols indicate the positions of the best-fit protoplanets.}
\label{fig:spiral_fit}
\end{figure}

We use Eq.~\ref{eq:spiral} to fit simultaneously spiral arms S1 and S2 in the $r^2$-scaled $Q_\phi$ $J$-band image \citep[e.g.,][]{muto2012,grady2013,benisty2015} by selecting 27 and 38 azimuthally equidistant points of surface brightness maxima along the S1 and S2 spiral arm, respectively. We aim to find two separate solutions by restricting the radial position of the protoplanets to be either inside or outside the scattered light cavity. The disk aspect ratio is limited to $0.05<h_{\rm p}<0.4$ such that it is consistent with radiative transfer modeling \citep{andrews2011,carmona2014} and the temperatures required are still physical. The rotational profile is assumed to be Keplerian with $\zeta=3/2$ and we use the flaring index, $\psi=\zeta-\eta-1=0.25$, from the radiative transfer model in Sect.~\ref{sec:radiative_transfer}. Changing $\eta$ did not significantly change the results so it was fixed to 0.25. The image is deprojected by the disk inclination after which we performed a $\chi^2$ minimization of the parameter grid which consists of 30, 70, and 70 linearly spaced values for $h_{\rm S1}$, $r_{\rm S1}/r_{\rm S2}$ and $\phi_{\rm S1}/\phi_{\rm S2}$, respectively. We fit the launching point of both spiral arms and the aspect ratio at the launching point of spiral arm S1. The aspect ratio at the launching point of the S2 spiral arm is scaled with the flaring index from the aspect ratio of the S1 spiral arm. We note that linear perturbation theory requires two protoplanets to explain the two spiral arms which will be discussed in Sect.~\ref{sec:origin_spiral_arms}.

The best-fit results, together with the parameter priors, are given in Table~\ref{table:spiral_best_fit} and the spiral arm solutions are superimposed on the $r^2$-scaled $J$-band image in Fig.~\ref{fig:spiral_fit}. For the protoplanet solutions inside the scattered light cavity, we are not able to fit the innermost points of the surface brightness maxima (see Fig.~\ref{fig:spiral_fit}). A better fit would require a disk aspect ratio larger than the upper limit of 0.4 which corresponds to temperatures that are unphysical at the disk radius that is studied. A similar result was obtained for the spiral arms of MWC~758 \citep{benisty2015}. The second scenario, in which the protoplanets are located beyond the scattered light cavity, gives a more realistic best-fit result for the aspect ratio, $h_{\rm S1}=0.16$, and the brightness maxima in Fig.~\ref{fig:spiral_fit} are traced well by the best-fit spiral arms. However, the radial and azimuthal coordinates of those protoplanets show a degenerate correlation because the launching point of the density wave (i.e., the kink in the spiral arm) is not found by the fitting routine. The 2D Bayesian probability distributions of each set of free parameters and their marginalized probability distributions are shown in Fig.~\ref{fig:prob_spiral} of Appendix~\ref{sec:spiral_probabilities}.

\section{Discussion}\label{sec:discussion}

\subsection{Shadows as a probe for the innermost disk regions}\label{sec:shadow_origin}

The polarized scattered light images that have been obtained with \mbox{SPHERE} show multiple local minima in surface brightness which we have interpreted as shadows that are cast by dust located in the vicinity of the star (see Sect.~\ref{sec:shadows}). Even though high-contrast imaging with \mbox{SPHERE} allows us to resolve disk structures at small angular separations from the star, the inner disk of HD~135344B lies at an expected angular separation of $\sim$7~mas which is well beyond the reach of \mbox{SPHERE}. Therefore, the shadow features and possible variability provide a unique way of probing the innermost disk regions.

Features A and B are narrow shadow lanes that could have been cast by a misaligned inner disk, similar to HD~142527 \citep{marino2015}, because their location appears to be stationary between epoch~1 and 2. A misaligned inner disk will be replenished by gas and dust flowing from the outer disk, through a disk warp, towards the inner disk \citep[e.g.,][]{casassus2015} with a mass accretion rate of $10^{-8}$~M$_\odot$~yr$^{-1}$ \citep{sitko2012}. A warped disk region might cast a (broad) shadow which is approximately bound by the shadows from the inner disk. This could explain shadow feature~D which manifests itself as a subtle dimming of the surface brightness.

Shadow feature~C is clearly detected in the $J$-band observation but appears to be absent in the $R$-, $I$, and $Y$-band observations which indicates a variable or transient origin of this shadow such as an instability in or perturbation of the inner disk \citep{sitko2012}, an accretion funnel flow \citep{muller2011}, an (aperiodic) outflow from the rapidly rotating star \citep{muller2011}, or an inclined circumplanetary disk.

To probe the dynamics of the innermost disk region, it would be possible to use variability of the casted shadows. In particular, an alternation in shape, location, and brightness of the shadow features gives an indication of the geometry, optical thickness, and variability timescale of the shadow casting dust. Time-dependent processes such as precession of a warped inner disk, asymmetric accretion flows, local perturbations of the inner disk, or a dusty disk wind may alter the shadows on various timescales.

Although a comparison of the two observation epochs shows a hint of variability for the direction and shape of shadows A and B (see Fig.~\ref{fig:polar_mapping}), the $R$-, $I$-, and $Y$-band observations are of significantly lower S/N than the $J$-band observations which makes it difficult to compare with high precision the shadow locations in the polarized scattered light images. In addition, the wavelength dependent dust opacities might alter the shape of the shadow features which could resemble variability. Future observations will determine if these differences are explained by changes in the inner most disk regions or by the wavelength dependent dust opacities. Also, follow-up observations may provide insight into the variable/transient nature of shadow feature~C which was only detected during in the second observation epoch.

A shadow that is cast by a clump in the vicinity of the central star will show an azimuthal gradient when the Keplerian timescale of that clump is similar to or smaller than the time delay related to the finite speed of light with which the shadow traverses the disk. The variable shadow feature~C shows a slight azimuthal gradient which could be a light travel time effect. The detected radial extent of shadow~C is approximately 0\ffarcs35 which translates into an 6.8~hr time of flight. Over this distance, the position angle of the shadow changes by $\sim$25$\degr$ which would correspond to the orbital rotation of the clump while the shadow is traversing the disk with the speed of light. Translating this into an orbital frequency gives a radius of $\sim$0.06~au.

\subsection{On the origin of the spiral arms}\label{sec:origin_spiral_arms}

Spiral arms have been observed in scattered light images of several protoplanetary disks, including HD~135344B \citep{muto2012,garufi2013,wahhaj2015}, MWC~758 \citep{grady2013,benisty2015}, HD~100546 \citep{boccaletti2013,avenhaus2014b}, HD~142527 \citep{casassus2012,rameau2012,canovas2013,avenhaus2014a}, and HD~100453 \citep{wagner2015}. They are caused by local perturbations of the disk surface density and/or scale height but their origin is still under debate. Interestingly, all these disks are in a transition stage and contain a large dust cavity which could be related to the observed spiral arms. Several mechanisms have been proposed, planet-disk interactions and gravitational instabilities will be discussed below.

Low-mass protoplanets embedded in a gaseous disk are known to excite density waves at Lindblad resonances which generates a one-armed spiral through their constructive interference \citep{goldreich1979,ogilvie2002}. Spiral arm fitting in Sect.~\ref{sec:spiral_arms} showed that the spiral arms are best explained by protoplanets located exterior of the spiral arms since this allows for a physical aspect ratio of the disk, in contrast to the protoplanet solutions interior of the spiral arms. However, it was assumed that each spiral arm is excited by an individual low-mass, $M_{\rm p}/M_* \lesssim 10^{-3}$, planet. Several authors have shown, using 3D hydrodynamical and radiative transfer simulations, that a massive, $M_{\rm p}/M_* \gtrsim 6\times10^{-3}$, protoplanet in the exterior of the disk will excite both the \mbox{m=1} and \mbox{m=2} spiral arm with a morphology, pitch angle, and surface brightness contrast similar to HD~135344B and MWC~758 \citep{dong2015,zhu2015,fung2015}. A kink is visible in the S1 spiral arm (see Fig.~\ref{fig:zimpol_color}) which could be related to the launching point of the spiral density wave by a forming protoplanet \citep{muto2012}. Fitting of the spiral arms did not result in a protoplanet solution at the location of the kink but Fig.~\ref{fig:spiral_fit} shows that the surface brightness maxima deviate from the best-fit solutions around this location.

\begin{figure}
\centering
\resizebox{\hsize}{!}{\includegraphics{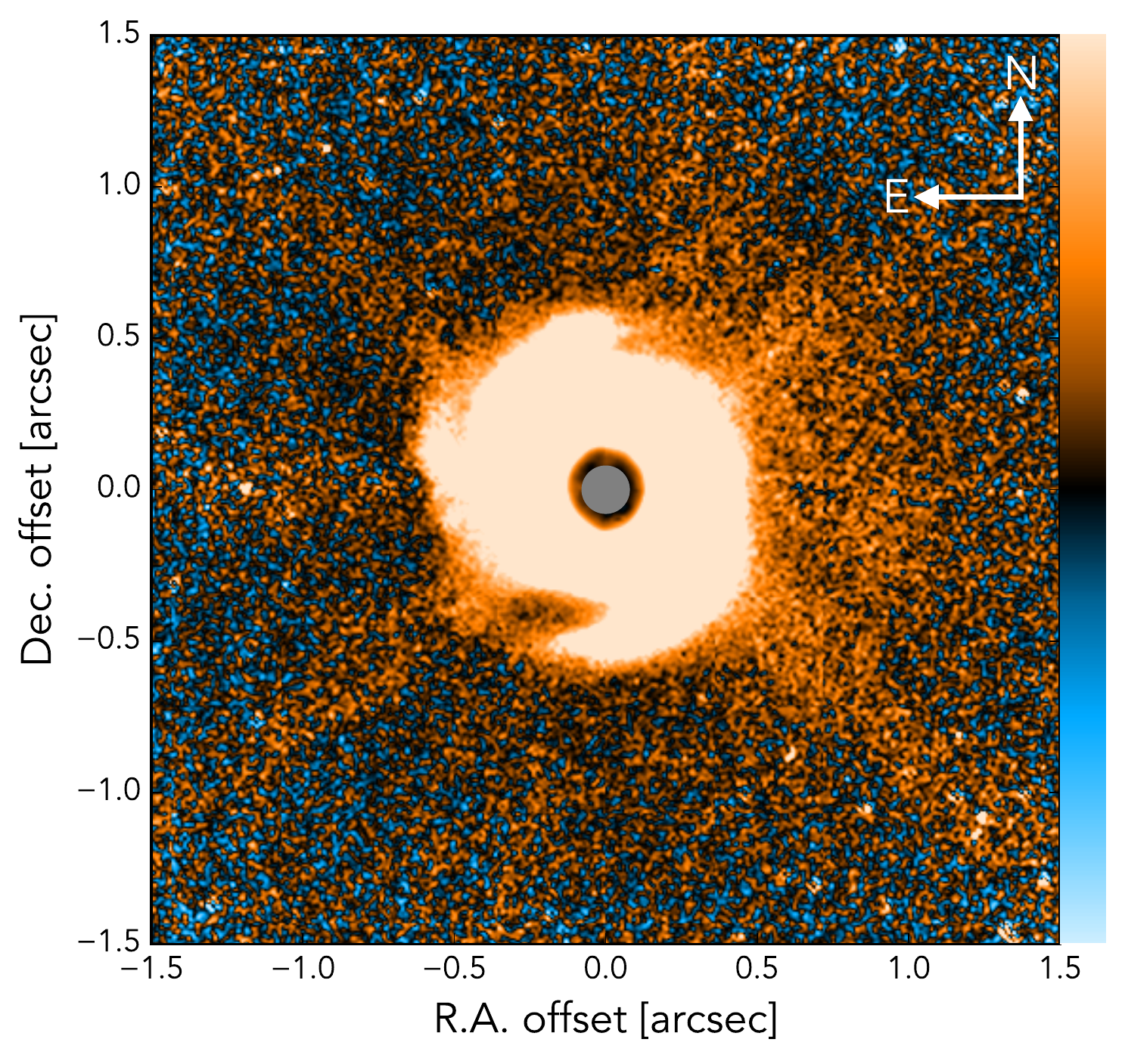}}
\caption{\mbox{IRDIS} $J$-band $r^2$-scaled $Q_\phi$ image shown with a ten times harder linear color stretch compared to the image in Fig.~\ref{fig:images_irdis}. Orange corresponds to positive pixel values, blue to negative pixel values, and black is the zero point. The 80~mas inner working angle of the coronagraph has been masked out. Features from the telescope spiders are clearly visible.}
\label{fig:hard_stretch}
\end{figure}

The gap width that will be opened by the gravitational torque of a forming protoplanet is approximately given by the Hill radius, $R_{\rm H}=r(q/3)^{1/3}$ with $r$ the disk radius and $q$ the planet-star mass ratio, but can be as wide as 5~R$_{\rm H}$ \citep{dodson2011}. A massive protoplanet exterior of the HD~135344B spiral arms is expected to have opened a wide gap which might have been visible in scattered light. Figure~\ref{fig:hard_stretch} shows the $J$-band $r^2$-scaled $Q_\phi$ image on a hard color stretch which reveals scattered light up to disk radii of 1\ffarcs0 without any indication of a gap exterior of the spiral arms. However, the depth of the gap in scattered light depends on the aspect ratio which, for a flaring disk, increases towards larger disk radii \citep{crida2006}.

Spiral density waves are also triggered by self-gravity in case the protoplanetary disk is massive enough to become gravitationally unstable \citep[e.g.,][]{cossins2009,kratter2010}. We use the Toomre parameter to estimate the stability of the HD~135344B disk against self-gravity which is given by \citep{toomre1964}
\begin{equation}\label{eq:toomre}
Q(r) = \frac{k(r)c_{\rm s}(r)}{\pi G \Sigma(r)},
\end{equation}
where $k(r)$ is the epicycle frequency, $c_{\rm s}(r)$ the sound speed, $G$ the gravitational constant, $\Sigma(r)$ the total surface density, and $Q(r)\lesssim 1$ corresponds to a gravitationally unstable disk. We assume a Keplerian disk, optically thin temperature profile and estimate the disk mass from the optically thin \mbox{ALMA} band~7 (0.85 mm) dust continuum \citep{pinilla2015}. The disk mass is given by
\begin{equation}\label{eq:disk_mass}
M_{\rm disk}(r) = \frac{F_\nu(r) d^2}{\kappa_\nu B_\nu(T)},
\end{equation}
where $F_{\nu}$ is the flux density, $d$ the distance to the source, $\kappa_\nu$ the millimeter dust opacity, and $B_\nu (T)$ the Planck function. We use $\kappa_\nu = 0.1(\nu/10^{12}{\rm Hz})$~cm$^2$~g$^{-1}$ \citep{beckwith1990} which implicitly includes a gas-to-dust ratio of $g/d=100$ and has a typical opacity slope for a protoplanetary disk in which grain growth has occurred \citep{testi2014}. However, we note that \citet{carmona2014} estimated $g/d=4$ from detailed modeling of line observations, which would imply a strong gas depletion throughout the disk. In contrast with \citet{vandermarel2016} who obtained $g/d=80$ from CO line and dust continuum observations with \mbox{ALMA}.

\begin{figure}
\centering
\resizebox{\hsize}{!}{\includegraphics{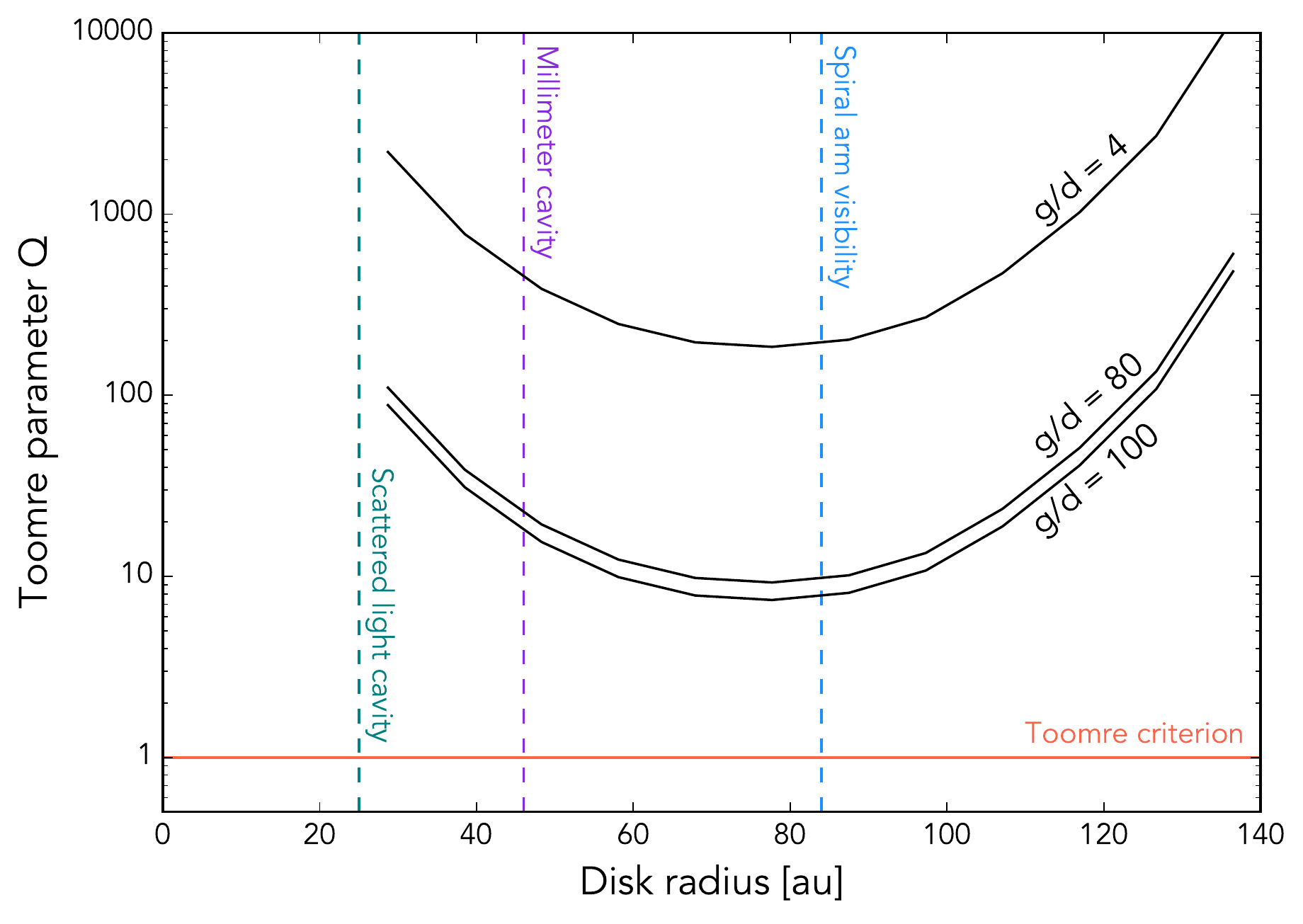}}
\caption{Toomre parameter calculated from the \mbox{ALMA} band~7 (0.85 mm) dust continuum \citep{pinilla2015} with a typical gas-to-dust ratio of $g/d=100$, $g/d=80$ \citep{vandermarel2016}, and $g/d=4$ \citep{carmona2014}. The dashed lines show the 25~au scattered light cavity, the 46~au millimeter cavity from \citet{andrews2011}, and the outer radius at which the spiral arms are visible.}
\label{fig:toomre}
\end{figure}

Figure~\ref{fig:toomre} shows the calculated Toomre parameter as function of disk radius with a minimum of $\sim$10 around 80~au which is an order of magnitude above the Toomre criterion when a gas-to-dust ratio of 100 is used. However, we note that the derivation of the Toomre parameter contains a number of uncertainties and assumptions such as the dust opacity, gas-to-dust ratio, temperature profile, and the exclusion of the dust mass in larger, centimeter sized grains. A marginal instability could have enhanced the amplitude and pitch angle of the spiral arms when triggered by one or multiple protoplanets \citep{pohl2015}.

\subsection{The interplay of surface and midplane}\label{sec:disk_midplane}

The polarimetric scattered light images show an overall positive surface brightness gradient towards the southwest side of the disk in the $r^2$-scaled $Q_\phi$ images of all filters. The gradient follows approximately the major axis and can therefore not be attributed to forward scattering. In addition, no strong forward scattering effect is expected given the low inclination of the disk (see the radiative transfer model in Sect.~\ref{sec:radiative_transfer}). Thus, the surface brightness gradient is presumably related to an asymmetry in the height of the $\tau=1$ disk surface. A local increase in scale height, surface density, or turbulence will alter the effective cross section and optical depth of the disk which results in a change in scattered light flux directed towards the observer. The surface brightness asymmetry is observed on a global scale up to 1\ffarcs0 (140~au) as shown in Fig.~\ref{fig:hard_stretch}. The integrated surface brightness is approximately a factor 1.1 larger on the west side of the disk compared to the east side, calculated by adding all $Q_\phi$ pixel values between 0\ffarcs7 and 1\ffarcs0 in two position angle bins from $45\degr$ to $135\degr$ and from $225\degr$ to $315\degr$ for east and west, respectively.

\begin{figure}
\centering
\resizebox{\hsize}{!}{\includegraphics{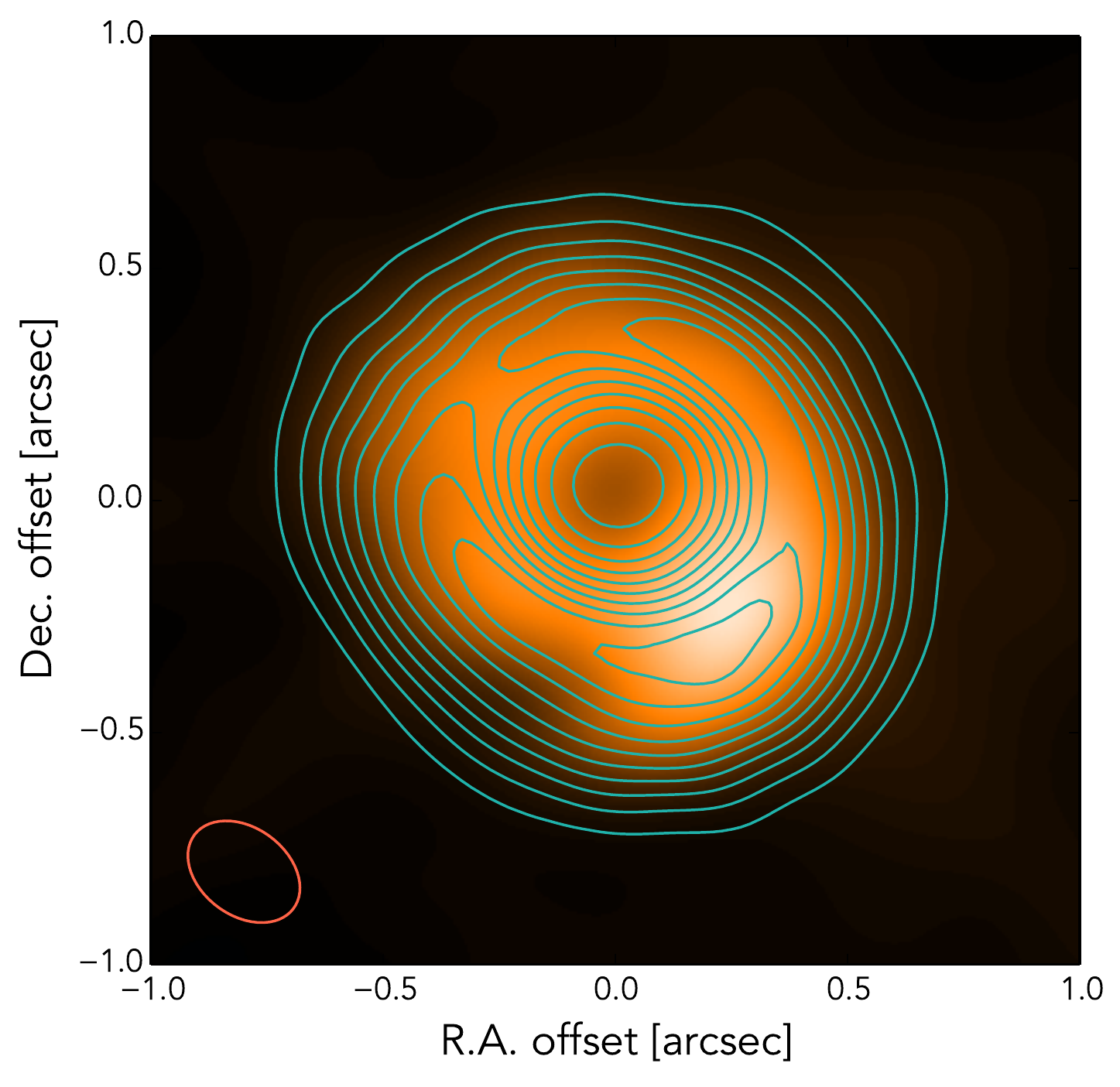}}
\caption{\mbox{IRDIS} $J$-band $r^2$-scaled $Q_\phi$ image convolved with an elliptical Gaussian kernel ($132 \times 94$~mas) to match the angular resolution of the \mbox{ALMA} Band~9 observation from \citet{perez2014}. The superimposed contours show the \mbox{ALMA} Band~9 (0.45 mm) dust continuum with the beam size in the bottom left. We assumed that the star is located in the cavity center of the \mbox{ALMA} image.}
\label{fig:sphere_alma}
\end{figure}

Figure~\ref{fig:sphere_alma} shows the $r^2$-scaled $J$-band $Q_\phi$ image convolved with an elliptical Gaussian beam ($132 \times 94$~mas) in order to match the angular resolution of the \mbox{ALMA} band~9 observation from \citet{perez2014} (contours in Fig.~\ref{fig:sphere_alma}). The brightness maxima of the scattered light and sub-millimeter observation approximately overlap which could be an indication that the asymmetry in the \mbox{ALMA} observations is related to the spiral arms in the PDI image \citep{perez2014}. More specifically, the crescent-shaped asymmetry might be resolved into spiral arms by employing the longest baselines available for \mbox{ALMA} \citep{quanz2015}. The offset between the peak of scattered light and millimeter emission could be a projection effect of a slightly inclined and flaring disk since millimeter observations probe the disk midplane whereas scattered light observations probe the disk surface.

Spiral arms excited by protoplanets possess an angular pattern velocity which is equal to the Keplerian angular velocity of the protoplanet. The gas surface density is expected to be larger within the spiral arms and shocks might be generated as the spiral arms traverse the disk which can locally affect the scattered light flux. The \mbox{ALMA} band~9 data (see Fig.~\ref{fig:sphere_alma}) is optically thick in the peak of the emission \citet{perez2014} and presumably traces the temperature of the disk. The peak emission coincides with a maximum of the $r^2$-scaled scattered light flux which might suggest that a temperature perturbation in the midplane has enhanced the disk height around the location of the kink in the S1 spiral arm. Future \mbox{ALMA} observations with higher angular resolution will reveal if the millimeter asymmetry is indeed related to the spiral arms and if the scattered light asymmetry can be explained by temperature and/or surface density perturbations in the midplane.

\section{Conclusions}\label{sec:conclusions}

We have carried out \mbox{VLT/SPHERE} polarimetric differential imaging observations in $R$- and $I$-band with \mbox{SPHERE/ZIMPOL} and $Y$- and $J$-band with \mbox{SPHERE/IRDIS} which reveal the transition disk around HD~135344B in scattered light with high angular resolution. We have studied the morphology and surface brightness of the disk in the context of the innermost disk regions, the origin of the spiral arms, \mbox{ALMA} dust continuum observations and dust grain properties in the disk surface. A possible interpretation of shadow features A, B, and D has been supported by an \mbox{MCMax3D} radiative transfer model. The main conclusions of this work have been summarized below.

\noindent Conclusions on the shadow features:
\begin{itemize}
    \item Multiple low surface brightness regions were discovered which have been interpreted as shadows casts by the innermost disk regions. An inner disk component that is $22\degr$ inclined with respect to the outer disk can explain shadow features~A and B. A warp in the disk, that connects the misaligned inner and outer disk, can explain shadow feature~D which is broad and bound by shadow features~A and B.
    \item Shadow feature C is clearly detected in the $J$-band observation but appears to be absent in the earlier $R$-, $I$-, and $Y$-band observations. The variable or transient nature of this shadow could be explained by several scenarios, including a local perturbation of the inner disk or an accretion funnel flow from the inner disk onto the star.
\end{itemize}

\noindent Conclusions on the spiral arms:
\begin{itemize}
  \item An explanation for the spiral arms could not be uniquely determined. In the context of linear perturbation theory, the spiral arms are best explained by two protoplanets orbiting exterior of the spiral arms. Protoplanet solutions inside the scattered light cavity seem unlikely because the spiral arm pitch angles would require unphysical disk temperatures.
  \item The surface brightness contrast and symmetry of the spiral arms indicate that a single massive protoplanet might have excited both the primary and secondary spiral arm interior of its orbit \citep{dong2015,zhu2015,fung2015}, however no gap is detected in scattered light beyond the spiral arms up to 1\ffarcs0.
  \item Alternatively, there could be a marginal gravitational instability around 80~au given the large uncertainties on the calculated Toomre parameter.
\end{itemize}

\noindent Additional conclusions:
\begin{itemize}
  \item The scattered light flux shows a positive gradient towards the southwest side of the disk in the stellar irradiation corrected $Q_\phi$ images which approximately coincides with the asymmetry of the sub-millimeter continuum emission. This could be the result of a temperature and/or surface density perturbation possibly related to the passing spiral arm.
  \item The disk color in polarized light is red which is an indication that large dust grains ($2\pi a\gtrsim\lambda$) dominate the scattering opacity in the disk surface. Large dust grains in the disk surface are expected to have an aggregate structure which provides them with aerodynamic support against settling towards the disk midplane.
  \item Part of the non-azimuthal polarization signal in the $U_\phi$ image of the $J$-band observation is likely the result of multiple scattering in the disk.
\end{itemize}

\begin{acknowledgements}

We are grateful to P.~Pinilla and C.W.~Ormel for useful discussions. We would like to thank L.M.~P\'{e}rez and P.~Pinilla for sharing their ALMA data and M.L.~Sitko for sharing the $R$- and $I$-band photometry data. We would like to thank the ESO staff and technical operators at the VLT for their excellent support during the observations. We are thankful to W.~Brandner for his valuable comments during the review by the SPHERE editorial board and we would like to thank the anonymous referee for a very thorough and detailed report, that helped us to improve the quality of this paper. Part of this work has been carried out within the frame of the National Centre for Competence in Research PlanetS supported by the Swiss National Science Foundation. S.P.Q., H.M.S., and M.R.M. acknowledge financial support from the SNFS. A.J. acknowledges the support of the DISCSIM project, grant agreement 341137 funded by the European Research Council under ERC-2013-ADG. H.A. acknowledges support from the Millennium Science Initiative (Chilean Ministry of Economy), through grant "Nucleus RC130007" and from FONDECYT grant 3150643. This research made use of Astropy, a community-developed core Python package for Astronomy \citep{astropy2013}. This research has made use of the SIMBAD database, operated at CDS, Strasbourg, France. SPHERE is an instrument designed and built by a consortium consisting of IPAG (Grenoble, France), MPIA (Heidelberg, Germany), LAM (Marseille, France), LESIA (Paris, France), Laboratoire Lagrange (Nice, France), INAF Osservatorio di Padova (Italy), Observatoire de Geneve (Switzerland), ETH Zurich (Switzerland), NOVA (Netherlands), ONERA (France) and ASTRON (Netherlands) in collaboration with ESO. SPHERE was funded by ESO, with additional contributions from CNRS (France), MPIA (Germany), INAF (Italy), FINES (Switzerland) and NOVA (Netherlands). SPHERE also received funding from the European Commission Sixth and Seventh Framework Programmes as part of the Optical Infrared Coordination Network for Astronomy (OPTICON) under grant number RII3-Ct-2004-001566 for FP6 (2004-2008), grant number 226604 for FP7 (2009-2012) and grant number 312430 for FP7 (2013-2016). This paper makes use of the following ALMA data: ADS/JAO.ALMA\#2012.1.00158.S and ADS/JAO.ALMA\#2011.0.00724.S. ALMA is a partnership of ESO (representing its member states), NSF (USA) and NINS (Japan), together with NRC (Canada), NSC and ASIAA (Taiwan), and KASI (Republic of Korea), in cooperation with the Republic of Chile. The Joint ALMA Observatory is operated by ESO, AUI/NRAO and NAOJ.

\end{acknowledgements}

\bibliographystyle{aa}
\bibliography{references}

\appendix

\section{Photometric calibration}\label{sec:photometric_calibration}

A photometric calibration is required for an absolute surface brightness comparison between filters. In this appendix, we will explain the photometric calibration procedure for the \mbox{ZIMPOL} and \mbox{IRDIS} observations.

The magnitude per pixel of the \mbox{ZIMPOL} observations is given by (Schmid et al., in prep.)
\begin{equation}\label{eq:zimpol_magnitude}
m^*(F) = -2.5\log\bar{C}^*(F) + zp(F) - X \ K_1(F) - m_{\rm mode}(F),
\end{equation}
where $\bar{C}^*$ is the mean count rate in each pixel, $zp(F)$ the zero point magnitude for a given filter $F$, $K_1(F)$ the atmospheric extinction parameter, $X$ the airmass of the observation, and $m_{\rm mode}$ a correction of the zero point for the instrument configuration. The zero point magnitudes for the $R$- and $I$-band filter are 24.29 and 23.55, respectively, the atmospheric extinction parameters are 0.106~mag~am$^{-1}$ and 0.078~mag~am$^{-1}$ for $R$- and $I$-band, respectively, and the instrumental mode correction for \mbox{\texttt{SlowPol}} mode of \mbox{ZIMPOL} is $m_{\rm mode}=-1.93$ (Schmid et al., in prep.). The airmass of the observations is provided in Table~\ref{table:summary_observations}. The pixel values of the $Q_\phi$ and $U_\phi$ images have been corrected for the DIT such that pixel values are given in counts s$^{-1}$. The magnitude per pixel is converted into a surface brightness with \citep{quanz2011}
\begin{equation}\label{eq:surface_brightness} 
S = m^* + 2.5\log{A},
\end{equation}
where $A$ is the square of the pixel scale in arcsec$^2$ and $S$ is in mag~arcsec$^{-2}$. The uncertainty in the \mbox{ZIMPOL} photometric calibration is approximately 10\%.

As a photometric check, we determine the magnitude of HD~135344B from the data and compare this with a literature value. We use the unsaturated frames that were obtained at the end of the \mbox{ZIMPOL} observations and correct those for both the DIT and the response function of the neutral density filter. We subtract 0.1~counts~s$^{-1}$ from each pixel to correct for the dark current level of a 10~s DIT with \mbox{ZIMPOL} as estimated from the PSF profile. We use circular aperture photometry (1\ffarcs5 radius aperture centered on the star) to determine the integrated flux and we assume this to be the photometric signal of HD~135344B. From Eq.~\ref{eq:zimpol_magnitude}, we obtain a magnitude of 8.48 and 8.12 for $R$- and $I$-band, respectively, which are very similar to the literature values given in Table~\ref{table:main_parameters}.

For the \mbox{IRDIS} observations, we do a more approximate photometric calibration as described in \citet{quanz2011} which leads to a larger photometric uncertainty than the \mbox{ZIMPOL} calibration. The $J$-band observation was carried out with a coronagraph but non-coronagraphic, unsaturated frames were obtained at the start and end of the observation (see Sect.~\ref{sec:observations}). The \mbox{IRDIS} $Y$-band observation was carried out without coronagraph and all frames are saturated in the PSF core. No unsaturated frames were obtained during the \mbox{HWP} experiment and we can not perform a photometric calibration on this data set.

We perform circular aperture photometry (1\ffarcs5 radius aperture centered on the star) on a reduced, unsaturated total intensity image in $J$-band and assume this to be the photometric signal from HD~135344B. Next, we correct the total intensity image, the $Q_\phi$ image, and the $U_\phi$ image for their DIT (see Table~\ref{table:summary_observations}). Also a correction is applied for the transmissivity of the $J$-band filter and the neutral density filter. We can now estimate the zero point of the \mbox{IRDIS} $J$-band observation by comparing the detector integrated $J$-band flux with the 2MASS magnitude. The zero point count rate is given by \citep{quanz2011}:
\begin{equation}\label{eq:zeropoint}
I_{\rm ZP} = I_{\rm CR} 10^{0.4m_{\rm 2MASS}},
\end{equation}
where $I_{\rm CR}$ is the aperture photometry count rate and $m_{\rm 2MASS}$ the 2MASS magnitude of HD~135344B. From the inversion of Eq.~\ref{eq:zeropoint}, we can convert the count rate for each pixel to magnitude and subsequently to surface brightness (see Eq.~\ref{eq:surface_brightness}). We note that the \mbox{IRDIS} photometric calibration includes a number of uncertainties and assumptions which leads to an estimated photometric error of 30-40\%.

\section{The cavity edge of the outer disk}\label{sec:outer_disk_cavity_edge}

\begin{figure}
\centering
\resizebox{\hsize}{!}{\includegraphics{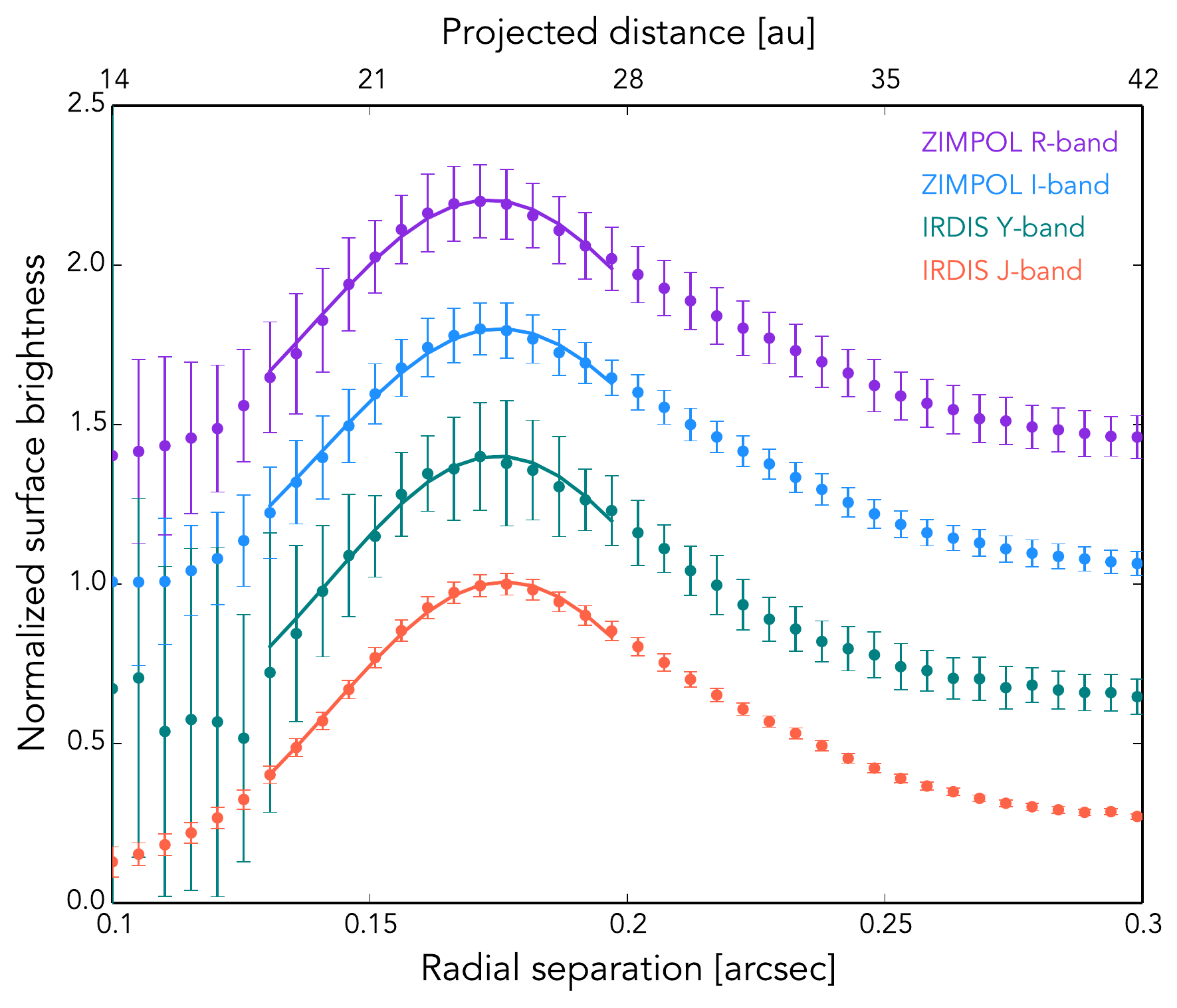}}
\caption{Azimuthally averaged polarized surface brightness around the cavity edge of the outer disk. The profiles are shown in normalized units and with arbitrary offsets. The data points show the mean $Q_\phi$ value in each bin and the error bars show 1$\sigma$ deviations in the corresponding $U_\phi$ bin. Solid lines show the best-fit Gaussian profiles (see also Table~\ref{table:disk_rim_results}).}
\label{fig:rim_profile}
\end{figure}

Spatial segregation of different dust grain sizes is expected to occur around a planet-induced gap edge \citep{zhu2012,ovelar2013}. The resolved millimeter cavity of the HD~135344B disk is indeed larger than the scattered light cavity as a result of dust filtration \citep{garufi2013}. We will investigate if dust segregation also affects the location of the outer disk cavity edge in the \mbox{ZIMPOL} and \mbox{IRDIS} images by calculating azimuthally averaged brightness profiles. The profiles of the $R$-, $I$-, $Y$-, and $J$-band images are shown in Fig.~\ref{fig:rim_profile} with each profile normalized to its peak-value and given an arbitrary offset. We perform a weighted least squares fit of a Gaussian profile to the points between 0\ffarcs1 and 0\ffarcs2 with the weights provided by 1$\sigma$ from the corresponding $U_\phi$ bins at the same disk radii. The location of the cavity edge is given by the brightness maximum and the width of the cavity edge is given by the FWHM of the Gaussian profile.

The best-fit values, provided in Table~\ref{table:disk_rim_results}, are very similar for all filters. The errors are derived from the $\chi^2$ minimization and are smaller than the spatial uncertainty from the pixel scales. We may conclude that there is no clear evidence of a wavelength dependence in the location and width of the cavity edge between the \mbox{ZIMPOL} and \mbox{IRDIS} data. The result is consistent with the red color of the scattered light (see Sect.~\ref{sec:disk_color}) since the scattering opacity in the optical and near-infrared will be gray for grains that are large compared to the wavelength. For smaller grains, the scattering opacity is larger at shorter wavelengths which would have caused a smaller cavity radius in the optical. FWHM of the cavity edge is approximately 80~mas which means that it is resolved with both the \mbox{ZIMPOL} and \mbox{IRDIS} observations (see Sect.~\ref{sec:observations}).

\begin{table}
\caption{Cavity edge best-fit results}
\label{table:disk_rim_results}
\centering
\bgroup
\def\arraystretch{1.25}
\begin{tabular}{L{2.6cm} C{2.4cm} C{2.4cm}}
\hline\hline
Filter & Maximum [au] & FWHM [au] \\
\hline
\mbox{ZIMPOL} $R$-band & $24.23\pm0.07$ & $11.33\pm0.28$ \\
\mbox{ZIMPOL} $I$-band & $24.55\pm0.06$ & $11.59\pm0.26$ \\
\mbox{IRDIS} $Y$-band & $24.47\pm0.08$ & $10.82\pm0.40$ \\
\mbox{IRDIS} $J$-band & $24.67\pm0.04$ & $11.11\pm0.14$ \\
\hline
\bottomrule
\end{tabular}
\egroup
\end{table}

\section{Spiral arm fitting probabilities}\label{sec:spiral_probabilities}

\begin{figure*}
\centering
\includegraphics[width=15cm]{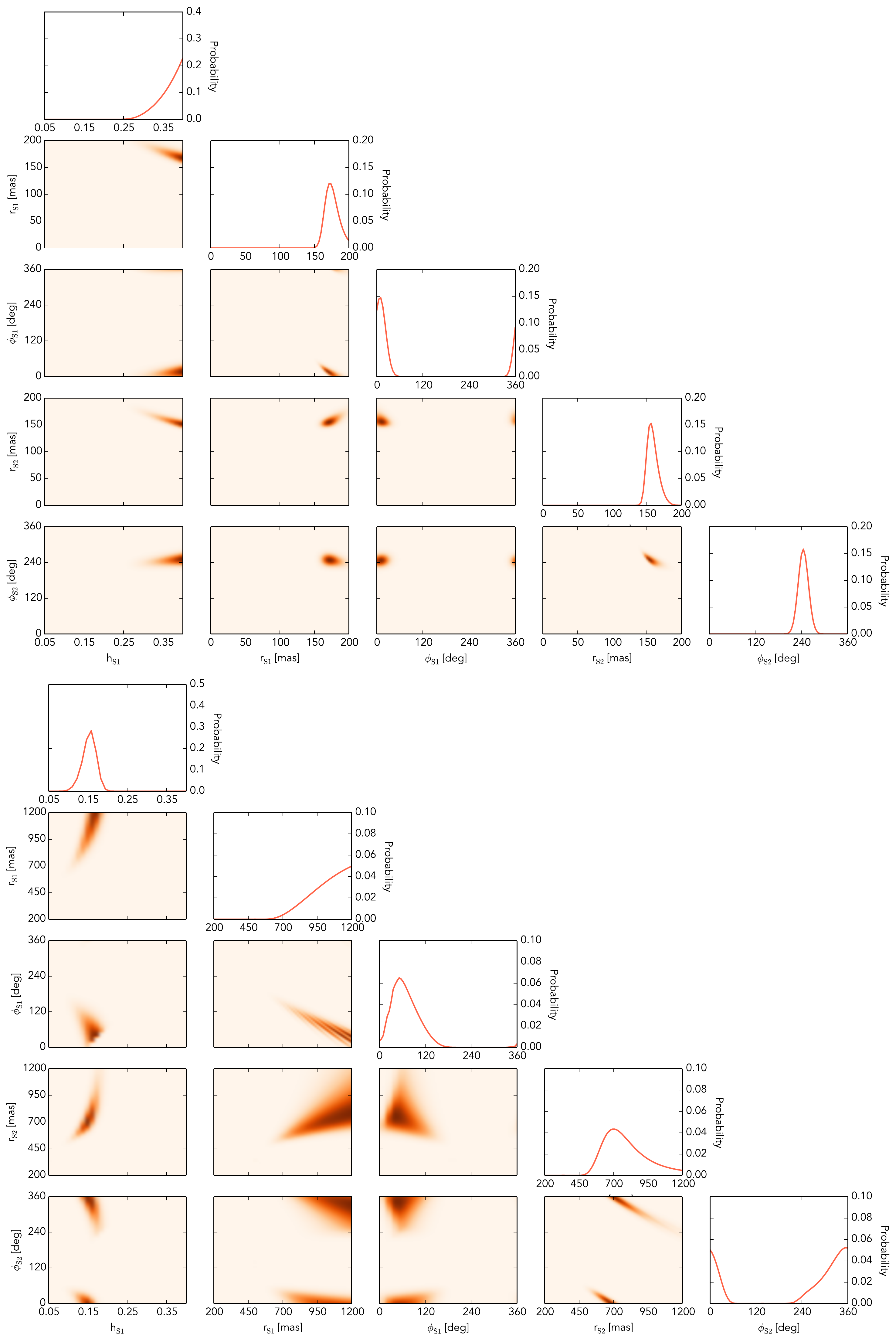}
\caption{Bayesian probabilities for the simultaneous fit of spiral arms S1 and S2 (see Fig.~\ref{fig:zimpol_color}). The protoplanet locations have been restricted to either inside (top triangle diagram) or outside the scattered light cavity (bottom triangle diagram). The exponent of the sound speed profile is kept fixed to $\eta=0.25$. The colored maps are the 2D probabilities of each set of free parameters and the top row of each column shows the marginalized probabilities of the free parameters that were fitted.}
\label{fig:prob_spiral}
\end{figure*}

Figure~\ref{fig:prob_spiral} shows the 2D Bayesian probability distributions of each set of free parameters that were used for the simultaneous fitting of spiral arm S1 and S2 as explained in Sect.~\ref{sec:spiral_arms}. The marginalized probabilities are shown in the top row of each column. The Bayesian probabilities are determined with $\exp{(-\chi^2/2)}$ from the $\chi^2$ solutions and normalized to a probability integrated value of unity.  

\end{document}